\documentclass[preprint2]{proto}
\usepackage{times}

\newcommand{\refs}{\par\noindent\hangindent=1pc\hangafter=1}
\newcommand{\eg}{\emph{e.g.},\ }

\begin{document}

\title{\textbf{\LARGE Modeling asteroid collisions and impact processes}}

\author {\textbf{\large Martin Jutzi}}
\affil{\small\em University of Bern}
\author {\textbf{\large Keith Holsapple}}
\affil{\small\em University of Washington}
\author {\textbf{\large Kai W\"unneman}}
\affil{\small\em Museum f\"ur Naturkunde}
\author {\textbf{\large Patrick Michel}}
\affil{\small\em Nice Observatory}

\begin{abstract}
\begin{list}{ } {\rightmargin 0.5in}
\baselineskip = 11pt
\parindent=1pc
{\small As a complement to experimental and theoretical approaches,  numerical modeling has become an important component to study asteroid collisions and impact processes.  In the last decade, there have been significant advances in both computational resources and numerical methods. We discuss the present state-of-the-art numerical methods and material models used in "shock physics codes" to simulate impacts and collisions and give some examples of those codes. Finally, recent modeling studies are presented, focussing on the effects of various material properties and target structures on the outcome of a collision.
 \\~\\~\\~}

\end{list}
\end{abstract}

\section{\textbf{INTRODUCTION}}

The modeling of impact processes can be based upon mathematical synthesis of experimental results, on direct theoretical application of the principles of physics, or on the use of those principles in numerical codes.   

The direct application of experimental results is not usually possible, because the experiments  cannot be performed at the actual conditions of interest. To bridge the gap, scaling theories are developed using physical principles to extrapolate experimental results to the actual conditions of interest. For some time the principal scaling theories have been based upon the physical concept of a "point source", wherein the genesis of an impact process is considered to occur instantaneously at a negligibly small region on the surface of the target object. Prior to 1982 the point source scaling was assumed to be governed by the kinetic energy of the impactor, but \emph{Holsapple and Schmidt} (1982) showed that such an assumption was not warranted, and extended the analysis to arbitrary point sources. Their general approach has been followed in numerous subsequent papers by \emph{Holsapple, Housen, and Schmidt}; one can refer to the review of those scaling approaches in \emph{Holsapple} (1993).  That scaling theory continues with applications to date.

In principle, the physics that governs such processes is known, at least in the continuum approach.  That physics includes the balance laws of mass, momentum, and energy, augmented by mathematical descriptions of the material behavior. Material behavior commonly includes the "equation of state", which models the hydrostatic components of the stress histories and the principal thermodynamics; as well as equations describing the deviatoric (shear) components of the response. The latter descriptions include stress-strain-temperature-relations and include also models for failure, flow, and fracture. That material behavior is the source of the primary uncertainties about the correct way to model these processes. However, there has been much progress in the last couple of decades, so that direct numerical solutions using time and space-stepping increments are becoming increasingly sophisticated and important.

The codes used in those numerical approaches are commonly called "hydrocodes", a remnant of the early days when they were used in the military industry to make calculations of weapons effects, and the modeling only included the hydrodynamic aspects of the processes. Nowadays those codes are better called "shock-physics codes" (\emph{Pierazzo et al.}, 2008).

Just within the last decade, another approach has been applied to asteroid processes. That approach has been borrowed from the fields of particle mechanics and of "n-body" studies.  They model the material of the body as a large number of individual, discrete, usually spherical, usually mono-sized, indestructible particles.  With those simplifications the balance of momentum alone determines the motions of the particles. Balance of mass is automatic, and there is no accounting for energy balance and heating. The interactions of the particles is modeled using combinations of concepts of restitution, friction, and most recently cohesive forces, with a number of interaction parameters.
These approaches have only been made possible because of the extensive growth of computing power. However, because of the extreme complications of the interactions of real particles at high energies and stresses, those approaches will most likely be restricted to cases where the stresses remain low, and the discrete particle nature of the process can identify processes not contained in the continuum approaches.  

In this chapter, we give an overview of the important asteroid properties that determine the outcome of a collision and discuss the physical processes involved. We then present recent experimental results and the theoretical approaches to describe the outcome of an impact. The main part of this chapter is devoted to a detailed discussion of the current state-of-the-art numerical models - shock physics codes - which are used in the field to simulate impacts and collisions. Some examples of those codes are presented and the various approaches to model the important properties and processes are detailed here, including hybrid hydrocode-particle code computations. Finally, some examples of recent modeling are presented. In these studies, the effects of various material properties and target structures on the outcome of a collision are discussed.

\bigskip
\centerline{\textbf{ 2. IMPORTANT PROPERTIES AND PROCESSES}}
\bigskip

Asteroids have complex shapes, internal structures and material properties. The impact response and mechanical behavior of such objects is naturally difficult to model. In this section, we present some of the important internal asteroid properties that determine the outcome of a collision and discuss late-stage processes. Asteroid interiors, morphologies and surface geophysics are further discussed in \emph{Scheers et al.}, \emph{Murdoch et al.} and \emph{Marchi et al. } (this volume).

\bigskip
\noindent
\textbf{ 2.1 Porosity}
\bigskip

 The fact that the bulk density of many asteroids is well below the grain density of their likely meteorite analogues indicates that many have significant porosity (\emph{Britt et al.,} 2002). In particular,  several lines of evidence point to the presence of a high degree of porosity for asteroids belonging to the C taxonomic class, as indicated by the very low bulk density ($\sim$ 1.3 g/cm$^3$) estimated for some of them, such as the Asteroid 253 Mathilde encountered by the NEAR Shoemaker spacecraft (\emph{Yeomans et al.,} 1997), and as inferred from meteorite analysis (\emph{Britt et al.,} 2006). That porosity might be a result of a rubble-pile structure, as suggested for instance by spacecraft observations of asteroid Itokawa (\emph{Fujiwara et al.} 2006). The various forms of structures and porosities of asteroids were discussed in Asteroids III (\emph{Britt et al.}, 2002; \emph{Richardson et al.}, 2002; \emph{Asphaug et al.}, 2002).
 
 It is useful to distinguish between "micro" and "macroporosity". The distinction is primarily a matter of scale. The terminology arose in the study of meteorites, wherein porosity not apparent to the naked eye was called microporosity, and visually obvious voids between a grain structure or other identifiable particles was called macroporosity. But in the context of numerical modeling those terms can take on different meanings, because there are three different ways to model void space. In the first way, which is appropriate when the void space is very small compared to any length scale of interest, the porosity is considered a continuum material property and modeled as part of the equations of state. That continuum approach to porosity modeling is discussed below. In the second way, some codes allow a single numerical cell to contain both material and void, and the resulting behavior is determined by a mixture theory. In both of these first two approaches, the porosity scale is smaller than a calculation cell.  In the third way, the void can be so large as to encompass an entire numerical zone, and zones without material can be scattered throughout a problem domain in some defined way.  The differences between the approaches will depend on the length scale of the modeled phenomena compared to the length scale of the porosity.

The response of an asteroid to an impact is strongly affected by the presence of porosity. In the outgoing shockwave, a porous material can undergo significant permanent compression and become hot, which creates a significant energy sink (\emph{Asphaug et al.}, this volume; \emph{Davison et al.}, 2010). That effect will be included in any numerical model where the size scale of the porosity is smaller then the width of the initial outgoing compression pulse.  Since it is typical in a code to numerically smear a shock over several calculation zones, even the third of the above porosity models can model significant crush up and energy loss, depending on the resolution of the calculation. 
In large scale collisions (say between bodies of a size of 100 km), a shock wave can lead to compression of porous bodies even if they contain large ($\sim$ kilometer size) voids, as long as these voids are smaller than the relevant scale (e.g. the impactor size). 

In addition to the effects at the shock, porosity and the material's resulting crushability can also have a dramatic affect on the entire cratering process.  Rather than an excavation processes, an entire crater can be formed by a downward flow crushing the material beneath the crater floor (\emph{Housen and Holsapple}, in preparation, 2014).

Recent experiments and the scaling theory for the regime of cratering dominated by target porosity will be discussed in section 3.
The different approaches proposed to model the various effects of porosity described above will be presented in section 4.4.5.

\bigskip
\noindent
\textbf{ 2.2 Strength}
\bigskip

The outcome of an impact into an asteroid, whether a crater or a disruption, will ultimately be determined by gravity and some strength measure of the material of the object. There are many measures of strength for a geological material, and, over the last decade or so, that variety has been identified and is often included in our scaling theories and in numerical calculations. Strength measures can include tensile strength, compressive strength, shear strength, crush strength and others; each governs the ability of a material to withstand a different kind of stress state.  In the usual continuum approach, each of those strengths is characterized by a different portion of a single "strength envelope": a boundary defined in stress space between elastic and inelastic (permanent) deformation. An additional part of the modeling (flow rules) then describes the nature of the inelastic deformation from flow or fracture. In addition, the prior shock history can modify or "damage" the material, and that also must be accounted for.  \emph{Holsapple} (2009) presented a review of strength theories appropriate for geological materials. The interested reader can refer to that reference, in addition to further detail below (section 4.4).

\bigskip
\noindent
\textbf{ 2.3 Late-stage processes}
\bigskip

Since the times of the early lunar studies it is been observed that the large craters and basins have a substantially lower depth to diameter ratio than the smaller ones. For the large craters, the slopes of the outer walls are typically well below the angle of repose expected for soils and rocks. These were judged to be puzzling because the angle of repose typically determines the static equilibrium slope angles of soils and rocks.
\emph{Melosh} (1979) proposed that the effect was due to the dynamic weakening of rock in the latter stages of crater formation by the action of acoustic vibrations.  Since then, many other calculators include a "late-time" period of crater formation using rheological models that suppose the presence of that acoustic fluidization.  Those methods are common today. An alternative viewpoint was presented by \emph{Holsapple} (2004a), but his approach has yet to be fully developed. 

These approaches are presented below in section 4.4.7.
An application of a dynamic weakening model in the case of large-scale collisions on Asteroid 4 Vesta (\emph{Jutzi et al.}, 2013) is discussed in the chapter by \emph{Asphaug et al.} (this volume).

\bigskip
\centerline{\textbf{ 3. SCALING LAWS}}
\bigskip

As mentioned in the introduction, the experiments we can make on Earth are not at the size scale, gravity levels, or impact velocities of interest to most of solar system impact events. For that reason, the results of experiments in the laboratory must be extrapolated, often over many decades, to predict the results of impacts into asteroids. How does a 10 km asteroid behave compared to a 10 cm lab sample?

The physical assumption forming the foundation of modern scaling theories is that of a general "point source". Any impactor has three fundamental independent measures: a radius $a$, a velocity $U$, and a mass density $\delta$.  Equally well the three independent measures can be taken as the diameter, momentum, and kinetic energy or any other three independent combinations. In any case, they contain the three independent units of length, mass and time.  Then, when that impactor collides at high velocity with an asteroid, it sets up a highly dynamic event affecting a region much larger than the impactor size, and over a timescale much longer than that of the initial deposition of energy into the surface. Physically, an appropriate assumption is that the deposition of momentum and energy is instantaneous into a region of vanishing dimensions compared to any length scale of interest such as the final crater.  Of course, that assumption cannot be made for very low speed impacts  or other cases where a final crater may be only slightly larger than the impactor.

For a point source, the governing impactor measures cannot retain a length scale or a timescale. From that assumption it follows that the individual values for the size, velocity, and mass density do not affect the outcome. Instead there can be at most one single combination of those 3 variables that "measures" the impactor. That measure is then used in conjunction with those defining the material behavior of the target object to develop the scaling theory. 

The earliest point source solutions for impacts simply assumed that the correct measure was the kinetic energy of the impactor, that defined what is now called "energy scaling". \emph{Dienes and Walsh} (1970) noticed in code calculations of impacts into metals that the same crater was obtained for different impacts having the same $ a U^\mu $ where $\mu=0.58 $. They did not identify that result as the signature of a point source, nor did they recognize that the general form is quite universal, but with different exponents depending on material type.  The Z-model of cratering by \emph{Maxwell} (1977) was essentially another early point source model, but again it was not identified as such, and it was only applied to the geometry of the cratering flow field.   \emph{Holsapple and Schmidt }(1982) developed crater scaling from the assumption of a general point source, measured by what they called a "coupling parameter". They showed that it must always have the power law form $C=a U^\mu \delta^\nu$, but the governing exponents $\mu$ and $\nu$ cannot be immediately predicted, because they depend in a complicated way on the material of the target. However, it was proved that $1/3<\mu<2/3$ (\emph{Holsapple and Schmidt}, 1982). They also applied the same measure to a variety of outcomes of cratering and determined definite interrelations between the power laws for all outcomes of a given event.

Over the years the implications of that theory have been applied to many different impact outcomes from both experiments and numerical simulations, including crater size, crater formation time, shock wave propagation time, catastrophic disruptions, ejecta characteristics from the impact, momentum transfer to asteroids, and others. From that variety of applications it is been well determined that the point source predictions work surprisingly well, and, for moderately porous materials $\mu \sim$ 0.4,  while for non-porous materials $\mu \sim$  0.55, with only small variations found.  And in many cases, the use of that assumption leads to simple power-law scaling for many features of interest. Well-known examples are the power laws for crater dimensions when a crater size is determined by the surface gravity ("gravity regime") or when it is entirely determined by a strength ("strength regime"), power laws for ejecta amounts and velocities, for stress decay, and others.

The reader interested in the details and the earlier applications might begin with the scaling review article in \emph{Holsapple} (1993). A review and several applications including the catastrophic disruption cases were presented in the \emph{Holsapple et al. }chapter in the 2002 Asteroids III volume.  

\begin{figure}[t!]
 \epsscale{0.9}
 \plotone{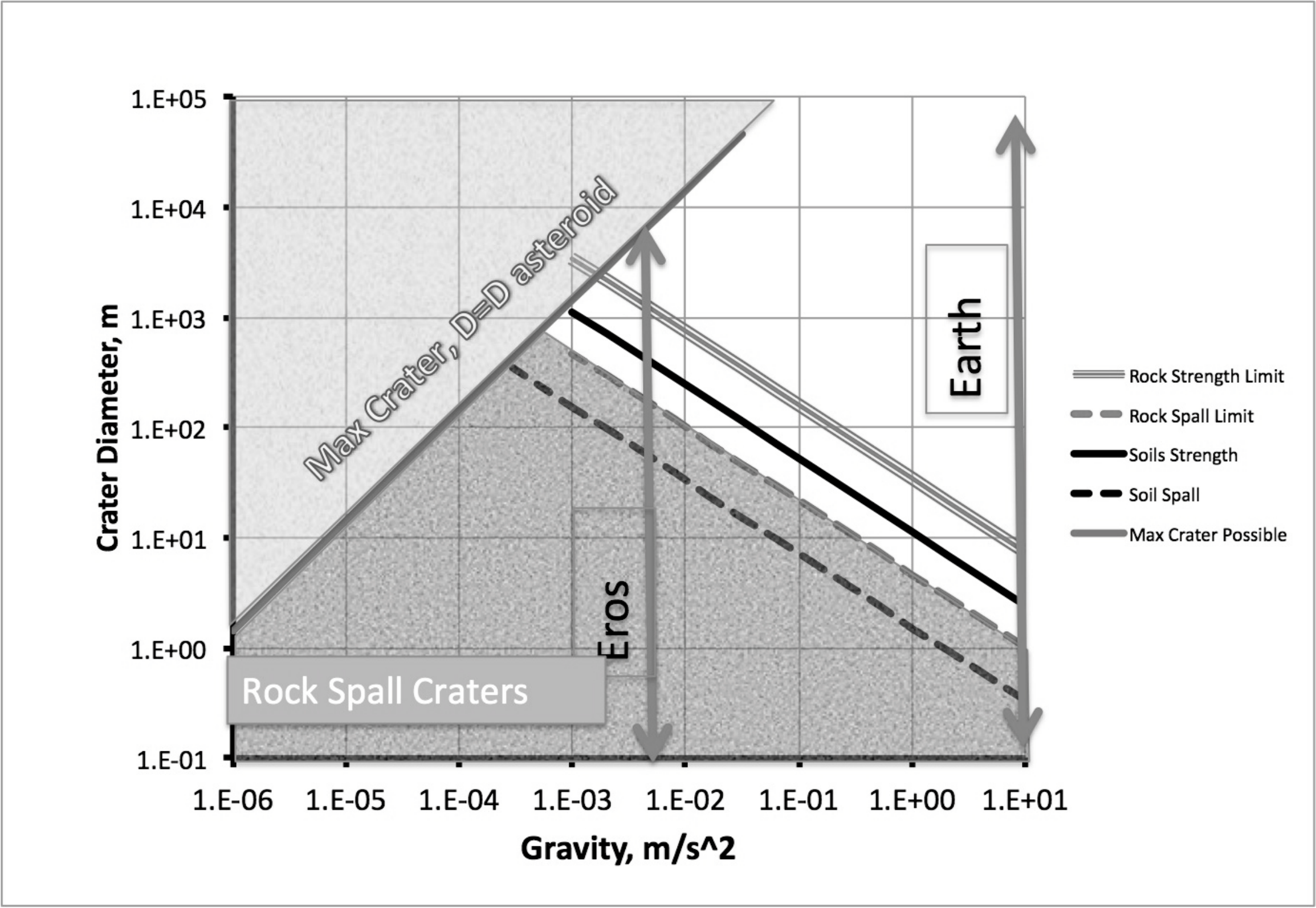}
 \caption{\small Cratering on rocky asteroids can be dominated by surface spall phenomena when gravity and the crater are small. The lower shaded region of this plot shows the extent of that region, it would include all craters possible on bodies of km size, and all craters smaller then 200 meters on Eros.}  
\label{figure1}
 \end{figure}

In addition to the now well-known strength and gravity regimes for impacts, two new regimes for cratering have been recently introduced.  First, \emph{Holsapple and Housen} (2013b) defined a ``spall'' cratering regime for small craters on rocky bodies, those are dominated by the tensile spall strength, and not the more common shear strength that determines excavation craters.  Those have been well known for explosive craters smaller than a meter or so in rocky targets on Earth and in laboratory experiments in competent rocks (e.g. \emph{Gault} 1973), although that regime has usually been ignored for planetary applications. But at the low gravity on an asteroid, that regime can include much larger craters. The extent of that regime as a function of surface gravity is depicted by the shaded region in Figure \ref{figure1}, from \emph{Holsapple and Housen} (2013b).  For a body such as Eros (16km) all craters smaller than about 1 km are predicted to be spall craters.  They would be flatter and shallower than excavation craters, and would eject blocks, not ejecta with more uniform smaller particles. These analyses are new and require more thorough experimental and numerical investigation.

\begin{figure}[h!]
 \epsscale{0.9}
 \plotone{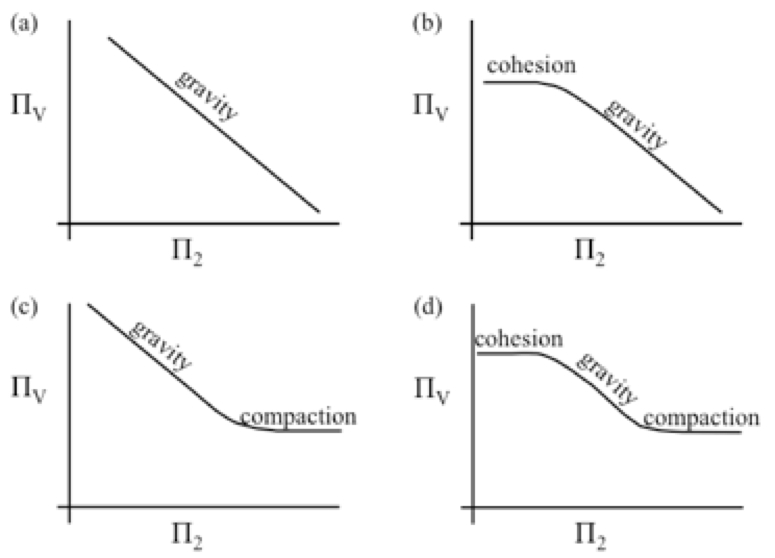}
 \caption{\small The scaled size $\Pi_V$ depends upon the gravity scaled size  $\Pi_2$ in different ways for different materials. (a) For a cohesionless soil with low or moderate porosity such as dry sand, standard "gravity scaling" applies at all size scales, and for increasing size or gravity there is a reduction in cratering efficiency.  (b) For a small or moderately porous material with cohesion, the cohesion is the dominant contribution to the shear strength at small crater size scales, so that the cratering efficiency is constant. But for increasing size, there is a transition to the gravity-dominated regime. (c) A cohesionless material with high porosity is gravity scaled at small sizes, but at large sizes is created by compaction, independent of gravity, so $\Pi_V$ approaches a horizontal asymptote. ( d) A material with both cohesion and high porosity potentially shows all three of the cohesion, gravity, and compaction regimes.}  
\label{figure2}
 \end{figure}

Secondly, \emph{Housen and Sweet} (2013) presented experiments and scaling of impacts into highly porous materials. Large porosity also adds a new regime, as illustrated in the schematic shown on a plot of scaled volume versus the gravity scaled $\pi 2$ parameter (Figure \ref{figure2}).  

As indicated above, the point source scaling theory has also been applied to catastrophic disruptions (e.g. since \emph{Housen and Holsapple}, 1990). Recently, \emph{Leinhardt and Stewart} (2012) extended this approach to define 'general' scaling laws that included collisions between gravity dominated bodies of comparable sizes. While this approach was shown to work well for some specific regimes (\emph{Leinhardt and Stewart} 2012), its general applicability still remains to be validated.

A great number of experimental studies in either the cratering or the disruptive regime have been performed since the publication of Asteroids III. This chapter is focused on the modeling of collisions; therefore we will not review this great amount of experimental work. We just give a few references to the interested reader noting that these experiments greatly contributed to our understanding of the collisional process at small laboratory scales and always give a necessary point of reference to test numerical methods (see Section 4.6).  In particular, various kinds of target materials have been considered such as targets made of sintered glass beads (e.g. \emph{Setoh et al.} 2010, \emph{Hiraoka et al.} 2011), pumice (\emph{Jutzi et al.} 2009), porous gypsum,  sometimes admixed with small pebbles or glass beads (\emph{Okamoto and Arakawa 2009}, \emph{Leliwa-Kopystynski and Arakawa} 2014, \emph{Yasui and Arakawa} 2011), dense (soda lime or quartz) cores and porous (gypsum) mantles (\emph{Okamoto and Arakawa }2008) and other porous materials (e.g. \emph{Housen and Sweet}, 2013; \emph{Nakamura et al.}, 2014). Moreover meteorites (e.g., \emph{Flynn et al.} 2008, \emph{Kimberley and Ramesh} 2011) have been used as well as 1-meter diameter granite spheres to check the effect of size on the cratering outcome (\emph{Walker et al.} 2013). In some cases, scaling parameters allowing an extrapolation at larger scales were also investigated. \emph{ Holsapple and Housen} (2012)  study experiments and scaling for the momentum coupling of impacts as related to the deflection of Earth-threatening asteroids by impacts (see also section 5.2.). A comprehensive review of experiments and ejecta scaling has been published by \emph{Housen and Holsapple} (2011).

The increasingly more sophisticated material modelling in shock physics code allows "numerical experiments" to be conducted for a much larger parameter space than it is possible to cover experimentally. Moreover the effect of individual properties such as friction and porosity can be investigated in detail. First results of this promising approach for the cratering regime have been presented in \emph{W\"unnemann et al.} (2010) and are summarized in section 5.1.1. Recent numerical studies of asteroid disruptions will be discussed in section 5.1.2.

\bigskip
\centerline{\textbf{ 4. NUMERICAL MODELING}}
\bigskip

\bigskip
\noindent
\textbf{ 4.1. Introduction}
\bigskip

As a complement to experimental and theoretical approaches, numerical modeling has become an important component to study the outcome of collisions.  In the last decade, advances in both computational resources and numerical methods have allowed the properties and processes involved (see section 2) to be modeled more and more realistically.

Here we present the principles of numerical impact modeling. We show some examples of codes used in the field and discuss the various model approaches.

\bigskip
\noindent
\textbf{ 4.2. Numerical techniques}
\bigskip

The two most common approaches to simulate impacts and collisions use shock physics codes ('hydrocodes') and particle codes.

Shock physics computer programs use continuum theory, and can calculate the entire dynamical processes including the propagation of shock waves and resulting fields of displacement, velocity, strain, stress, etc., as function of time and position (e.g. {\em Anderson}, 1987). They rely on mathematical constitutive models: for the thermodynamics which often includes melt and vaporization, for the deformation processes, and for the failure, fracture and flow. Those material equations are the outcomes of testing of materials at the various states of interest. The equations are solved in a time-stepping manner on a geometrical computational grid (or interpolation points), usually with zones much smaller that the impactor dimension. The allowable time step size must be shorter than the time for the passage of a shock wave across the smallest zone ("Courant-Friedrichs-Levy" stability criterion). Depending on the numerical method used, additional time step restrictions are required (e.g. {\em Anderson}, 1987). It is not uncommon in 2D applications to include hundreds ($n$) of space zones in each direction, for a total of $n^2 \sim 10^5$ and runs for several $10^6$ time steps. In 3D problems, many more zones ($n^3$) are required, so generally $n$ must be much smaller. That illustrates the great advantage of codes that can calculate in 2D. However, many problems are inherently 3D.

In contrast, particle codes (including "discrete element codes") assume a collection of simple interacting particles.  These codes generally only perform 3D calculations.  Their collisions are modeled using heuristic descriptions for restitution, friction and viscosity, and their interactions include mutual inter-particle gravitational forces. The balance of linear and angular momentum is all that is required to calculate their motions and rotations, so there is no use of mass balance or energy balance concepts.  They do not include crushing, melting, vaporization or other phenomena occurring in high-speed impacts, so these models are limited to low velocity, low stress events.  The number of particles might be millions, but that is still many orders of magnitude less than the actual number in a soil-like structure in any problem of interest.  Thus, it is inherently assumed that there are "enough" particles to approach the "infinite-number", continuum limit. And, to date, it appears that the governing parameters need be chosen on a case-by-case basis.

There are two classes of particle models. The simpler and earlier approaches are the so-called "hard sphere" models in which particle collisions, which are assumed to occur instantaneously, are predicted in advance and are governed entirely by coefficients of restitution chosen by the user. Such a model has been implemented in the N-body hard-sphere discrete element code \emph{pkdgrav} (\emph{Richardson et al.}, 2000), which has been used for different applications in Planetary Science (see \emph{Michel et al. }this volume for an application to asteroid family formation) and which has been adapted to enable dynamic modelling of granular materials in the presence of a variety of boundary conditions (\emph{Richardson et al.}, 2011). Other examples of such codes are polyhedral rubble piles codes (\emph{Korycansky and Asphaug }2006; \emph{Movshovitz et al. }2012). 

The hard sphere approach is equivalent to the early attempts to model fluid mechanics by a collection of very sparse interacting atoms or molecules. It works well for rarefied gas dynamics and it is well known that the averaging across many particles leads to the classical continuum equations of perfect  fluids. However, dense systems involving multiple collisions and enduring contacts require another approach, such as the "soft sphere" discrete element method. In this approach, the particles are allowed to have finite interaction times governed by elastic concepts, and in principle even static cases of enduring contact can be included relying upon modest penetration between particles. The soft sphere approach was recently included in the code \emph{pkdgrav} (\emph{Schwartz et al.,} 2012); it is also used in other codes (e.g. \emph{Sanchez and Scheers}, 2011). Moreover, these discrete approaches can include bonding forces between particles, as a first analog of cohesive materials (see, e.g., \emph{Schwartz et al.,} 2013, \emph{Sanchez and Scheeres}, 2014).  Such capability is also now available in the commercial finite-element code, LS-Dyna.  

But this soft-sphere extension generally comes at a price. In effect, although the soft sphere approach has the advantage of not requiring collisions to be predicted in advance, it comes at the expense of much smaller integration timesteps than the hard sphere approach, which can limit the integration timescale. That is similar to the time step restrictions for the continuum codes. On the other hand, because it can be implemented into codes, like \emph{pkdgrav}, that are fully and efficiently parallelized, it is currently possible to follow the evolution of millions of particles over a fairly large range of conditions. \emph{Murdoch et al.} (this volume) give a review of discrete element methods and continuum approaches applied to the dynamics of granular materials at the surface of asteroids.

The use of particle codes to simulate rubble pile collisions at low speeds (below the sound speed of the materials) was discussed in Asteroids III (\emph{Richardson et al.} 2002). In a recent study, \emph{Ballouz et al.} (2014a, b) used the particle code \emph{pkdgrav} to simulate low-velocity collisions between rotating rubble pile in order to measure the effect of the initial rotation of colliding bodies on the outcome. 

In situations where the whole process of a large-scale asteroid collision is investigated, including both the initial shock passages, heating, fragmentation and the subsequent reaccumulation of fragments, a hybrid approach is often used (e.g., \emph{Michel el al.}, 2001, \emph{Michel et al.}, this volume). There the fragmentation is computed with a shock physics code and the gravitational reaccumulation with a particle code. This approach is limited by the numerical resolution that fixes the minimum size of tractable fragments (typically down to $\sim$ 10 meters for simulations involving km-sized bodies).

In the fragmentation phase of a hypervelocity asteroid collision, self-gravity can usually be neglected\footnote{However, for large asteroids it may be important to include fracture shielding due to compression} because 1) the overburden pressure is small compared to the amplitude of the shock wave and 2) the fragmentation timescale is much smaller than the reaccumulation timescale. The fragmentation timescale is given by the time it takes for a shock wave to travel trough the whole target $\tau_f ~ \sim R_t/c_s$ where $R_t$ is the target radius and $c_s$ a wave speed.  On the other hand, gravitational reaccumulation proceeds on a time scale of $\tau_{dyn} ~ \sim (G\rho)^{-1/2} \sim$ 2200 s (for a bulk density $\rho$ = 3000 kg/m$^3$ and  the gravitational constant $G=6.67\times 10^{-11} m^3 kg^{-1} s^{-2}$). For small bodies ($R_t \lesssim$ a few 100 km), $\tau_f \ll \tau_{dyn}$ and gravity does not affect the dynamics of the fragments during the fragmentation phase. However, it is important to note that for very low velocity collisions, gravity has to be computed during the whole process, even for small bodies. This is typically the case in accretionary collisions (see \emph{Asphaug et al.}, this volume) where the impact velocity $v_{imp}$ is of the order of the mutual escape velocity $v_{esc} = \sqrt{2 G (M_p+M_t)/(R_p+R_t)}$,   where $M_p$ is the mass of the projectile, $M_t$ the mass of the target and $R_p$ the projectile radius. For 10 (100) km diameter bodies, $v_{esc} \sim$ 5 (50) m/s (assuming a density of $\rho$ = 3000 kg/m$^3$).

The hybrid hydrocode-particle code approach is detailed in \emph{Michel et al.} (this volume), where also the newest collision models in particle codes are presented. Here we focus on shock physics codes. 

\bigskip
\noindent
\textbf{ 4.3. Basic equations}
\bigskip

Shock physics codes  solve the system of partial differential equations that describe the conservation of mass, momentum and energy for a continuous, compressible medium (see \eg \emph{Collins et al.}, 2013 for a recent review). Examples and a description of such codes used in the field (SPH, CTH, iSALE, SOVA) will be given in section 4.5.

The stress tensor is symmetric and is often divided into isotropic (hydrostatic) and deviatoric parts
\begin{equation}
\sigma^{ij}=S^{ij}-P\delta^{ij}
\end{equation}
where the pressure $P$ is the hydrostatic pressure, $S^{ij}$ is the (traceless) deviatoric stress tensor and $\delta^{ij}$ is the Kroneker symbol.
Using a Lagrangian reference frame, the conservation equations (mass, momentum and internal specific energy) can then be written using an indicial summation convention as follows:
\begin{equation}
   \frac{d\rho}{dt}+\rho\frac{\partial v^{i}}{\partial x^{i}}=0
   \label{eq:massconv}
\end{equation}

\begin{equation}
   \frac{dv^{i}}{dt}=\frac{1}{\rho}\frac{\partial\sigma^{ij}}{\partial x^{j}} + g^i
      \label{eq:momconv}
\end{equation}
\begin{equation}
\frac{dE}{dt}=-\frac{P}{\rho}\frac{\partial}{\partial x^{i}}v^{i}+\frac{1}{\rho} S^{ij}
\bar\epsilon^{ij}
\end{equation}
where $d/dt$ is the Lagrangian time derivative, $\rho$ the density, $v$ the velocity, $E$ the specific internal energy, $x$ the position and
 $\bar\epsilon$ is the deviatoric part of the strain rate tensor. The term $g$ on the right hand side of eq. (\ref{eq:momconv}) accounts for any external acceleration, for instance due to gravity forces.

To complete the set of equations, equations describing the material response are required.  For the hydrostate components an equation of state (EOS) is required, which relates pressure, density and internal energy (or temperature, if its inclusion is convenient). And constitutive equations relating the deviatoric components are used.  Both of these can be very complex, but have been developed over the years, often by the military community.  The specifics are discussed in section 4.4. 

For many problems, the acceleration due to self-gravity is important and has to be taken into account in eq. (\ref{eq:momconv}). The components of the gravity acceleration can be computed by solving the Poisson equation for the gravitational potential. In particle based approaches (including SPH), the gravity acceleration for a particle $k$ is directly given by
\begin{equation}
\vec g_k = -G \sum_{k \ne l} \frac{m_l}{r_{kl}^2}\frac{\vec r_{kl}}{r_{kl}}
\end{equation}
where $G$ is the gravity constant, $m$ the mass of the particles and $r=|\vec r|$ the distance between the particles. For $N$ particles, the direct summation method leads to a complexity of O($N^2$) and is only practical when computers which are designed to solve such problems (e.g. GPUs) are used. A method often applied for the self-gravity computation is the Barnes-Hut tree algorithm (\emph{Barnes and Hut}, 1986) which allows to  reduce the complexity to O($N$ log $N$).


\bigskip
\noindent
\textbf{ 4.4. Material models}
\bigskip

An equation of state (EOS) relates density, internal energy and pressure, and may also include porosity effects. A stiffness and strength model is needed to determine the deviatoric stress due to strains and possible material failure (section 4.4.2 and 4.4.3).

Various processes occur during an impact on an asteroid and these have to modeled in a suitable way. Those may include the following: 

\begin{itemize}
\item effects due to porosity and other non-linearities
\begin{itemize}
   \item energy dissipation by compaction
   \item damping of shock wave
   \item reduction of wave speed in highly porous materials
\end{itemize}
\item (post-impact) flow of granular material in the case of shattered asteroids or rubble piles
\item dynamical state weakening (e.g. important to model the collapse of large craters)
\end{itemize}

\bigskip
\noindent
\textit{ 4.4.1. Equations of state}
\bigskip

An often used form for the EOS is 
\begin{equation}
P=P(\rho,E)
\end{equation}
This form is convenient because in most hydrocodes, the specific internal energy (rather than the temperature) is computed directly. In fact, temperature is not required to solve the conservation equations.  However, it may be useful to account for phase transitions and the thermal softening when deviatoric stresses are included (section 4.4.2). 

The most simple equations of state have no thermodynamic coupling and the pressure is solely a function of density (e.g. the Murnaghan EOS). A  more sophisticated and widely used analytical equation is the Tillotson EOS, which was  derived for high-speed impact computations (\emph{Tillotson}, 1962). One of the major advantages of this EOS is its efficiency. However, the Tillotson equation of state does not provide information about how to compute the temperature or the entropy of a material. Furthermore, the treatment of vaporization is not very sophisticated. For these reasons, the Tillotson EOS is mostly used to study impacts involving specific energies which do not lead to significant melting or vaporization. 

A more complex and thermodynamically consistent analytical EOS model is ANEOS (M-ANEOS) (\emph{Thompson and Lauson}, 1972; \emph{Melosh} 2007). In this model, the thermodynamics variables are derived from the Helmholtz free energy. ANEOS includes a more accurate treatment of both melting and vaporization than the Tillotson approach and allows for other polymorphic and liquid/solid phase transitions. 

There are also complete tabular databases such as the SESAME library, developed by the Livermore National Laboratory. Those are efficient to use, valid over vast density ranges, and commonly include complete thermodynamics including phase changes of melt and vapor. Often those are tabular compilations of the analytical forms.

\bigskip
\noindent
\textit{ 4.4.2. Strength models}
\bigskip

The strength model is fundamental for modeling impacts and collisions involving small bodies.   It determines the 'impact strength' in disruptive collisions, the size, final shape and characteristics of the crater in an impact, and so on. 

It is important to note that, depending on the loading conditions, various forms of 'strength' exist (section 2). Here we give an overview of some strength models that are included in impact codes used in the field. 

A simple strength model commonly used is the von Mises model developed for ductile metals. In this model, plastic flow occurs at stresses larger than a single constant yield strength $Y_0$. The von Mises criterion is implemented in shock codes by reducing the deviator stress (see. e.g. \emph{Benz and Asphaug }1994, 1995) by:
\begin{equation}
S^{ij} \rightarrow f S^{ij}
\end{equation}
where $f$ is computed by
\begin{equation}
f = \mbox{min}\left[\frac{Y_0^2}{3 J_2},1 \right]
\end{equation}
This is commonly called the "radial return method" and is also, for the von Mises case, the direction of the "associated flow rule" of plasticity theories. Here, the second invariant of the deviatoric stress tensor $J_2= \frac{1}{2} S^{ij} S^{ij} $ is used as a scalar measure of the maximal shear stress.

Although this model was developed for ductile materials, it was commonly used in impact calculations in geological materials in the past. In combination with a tensile fracture model (see section 4.4.3) it gives reasonably accurate results in disruptive collisions. \emph{Jutzi} (2014) gives a comparison between this approach and more sophisticated strength models in simulations of disruptive collisions. 

However, common geological materials such as soils, rocks and ices have more complex behavior than ductile metals. An important characteristic of their strength is a substantial  increase of shear strength with increasing confining pressure. That is the common feature of more sophisticated failure criterion used for geologic materials. The general form for geological strength models is as indicated in the Fig. 1 of \emph{Holsapple} (2009). There is a region in "shear-pressure" space delimited by a closed curve at which "failure" or "flow" can occur.  The "shear" can either be measured by the maximum shear stress $\tau$ on any plane, or by the average shear given by the stress invariant $\sqrt{J_2}$.  The "pressure" can as well be the actual average normal stress $P$ or the maximum normal stress $\sigma$ on any plane.  The Drucker-Prager model uses $P$ and $\sqrt{J_2}$, while the Mohr-Coulomb model uses $\sigma$ and $\tau$.  For small pressure, failure occurs at the upper limit pressure-dependent shear envelope defining the largest admissible shear. For large pressures, that shear envelope flattens.  Then, at the right is a constraint for the maximum pressure where compaction can occur with compressive pressure.  This limit is particularly important for porous material for which "crushing" can start at very low pressures. Models for that right-most porous limit are described below.

The initial slope of the shear envelope at low or tensile pressures is commonly called the "friction coefficient", because the form is similar to that used for friction between solid sliding blocks. But, in fact that is a misnomer, the actual mechanism is a result of the fact that a pressure impedes the movement of irregular-shaped grains up and over each other in a shearing flow.  That slope then reduces at larger pressures.

The Mohr-Coulomb and Drucker-Prager models include that pressure dependent yield strength and are the simplest failure models commonly used in soil and rock mechanics (e.g. \emph{Holsapple}, 2009).

One strength model (\emph{Collins et al.}, 2004) uses the following pressure dependent shear strength for the intact material $Y_i$:
\begin{equation}
Y_i = Y_0+ \frac{\mu_i P}{1+\mu_i P/(Y_M-Y_0)},
\label{eq:Yi}
\end{equation}
where $Y_0$ is the shear strength at $P$ = 0 and $Y_M$ is the shear strength at $P$ = $\infty$ and $\mu_i$ is related to the coefficient of internal friction for the intact material. Here the "shear" is measured by $\sqrt{J_2}$. This model is implemented in a number of codes (e.g. \emph{Collins et al.}, 2004; \emph{Senft and Stewart}, 2007; \emph{Jutzi}, 2014). Other nonlinear functions including powers and exponentials are common (e.g. \emph{Hoek and Brown,} 1980).

To describe the yield strength of fully damaged rock, which includes granular material,
\begin{equation}
Y_d = \mu_d P
\label{eq:Yd}
\end{equation}
is used\footnote{For modelling regoliths in low gravity environments, it is not uncommon to add a (small) cohesion}, where $\mu_d$ is related to the coefficient of friction of the damaged material.

In addition, the model must specify how a stress exceeding the limit is mapped back to the failure surface in each time step (the "return method") and also the resulting plastic flow increment during the time step (the "flow rule").   The yield strength $Y$ is often used to simply reduce the deviatoric stress (represented by $\sqrt{J_2}$) by a factor of $Y/\sqrt{J_2}$. That is again the "radial return method".  The plastic strain increment is often assumed to be in shear only (no volume change).  But in general plasticity theories, the "associated flow rule" is more common, it is assumed to be in the direction perpendicular to the failure envelope. That case includes the important phenomena of the dilantancy of a granular material when flowing.  Such details are not yet included in most planetary code simulations, although such an approach has been implemented recently in the iSALE code (\emph{Collins} 2014). Much more study and use of such models is warranted.

One might note the occurrence of two characteristic strength measures in such equations. The strength $Y_0$ at zero pressure is commonly called the cohesion. It may be zero for dry sands, or on the order of a few kPa to several MPa or more for cohesive materials, and even hundreds of MPa for small solid rocks. Then the shear strength increases with pressure to a maximum value of $Y_M$, a characteristic of the strength of individual grains, which may be as much as a few GPa.  The importance of these values depends on the other pressure scales in the problem, and especially on the gravity-induced lithostatic stress $\rho g h$ at a depth $h$.  If that gravitational stress is much larger than the cohesion, but still less than $Y_M$, then the cohesion can be ignored, but a dependence on the angle of friction will still occur.  That is the case in what is called the "gravity regime" of cratering which holds for instance for large craters on Earth. In the gravity regime, there remains a dependence on that angle of friction, but not on any other strength measure.  That dependence will emerge from any calculation including the strength equation.  But for extremely large events, the gravity stress will be larger even than $Y_M$.  For example, for collisions between planetary-sized bodies, that will be the case. Therefore, in giant impact simulations, for example to study the Moon forming impact (e.g. \emph{Benz et al.,} 1989; \emph{Canup and Asphaug}, 2001; \emph{Reufer et al.}, 2012), any form of strength is typically ignored. 

There are additional factors which must be accounted for in the construction of any strength envelope. It is not static, but changes according to the present state of the material at any point. For example, it can depend on the instantaneous strain rate, or on the temperature, and can change size and shape as the material is strained. The community is just beginning to include all known effects in our mathematical models.

For example, to account for the thermal softening of geologic materials for extreme impacts, $Y$ may be further reduced according to:
\begin{equation}
Y \rightarrow Y \tanh\left\{ \xi \left(\frac{T_{melt}}{T}-1\right)\right\}
\end{equation}
(Collins et al., 2004), or using an similar function of specific internal energy $E$: 
\begin{equation}
Y \rightarrow Y \left(1-\frac{E}{E_{melt}}\right)
\end{equation}
where $E$ is the specific internal energy and $E_{melt}$ the specific melting energy. It is important to note that the temperature and energy at which melt occurs are strongly pressure dependent and not simply a constant, and that fact should be included in any thermal softening equation. A possible approach is to use the location of the present thermodynamic state compared to the melt boundary in the EOS space to soften the material.

\bigskip
\noindent
\textit{ 4.4.3. Fracture (or damage) models}
\bigskip

There is an additional complexity when a rock fractures.  When a rock is stressed to failure, it breaks and becomes granular; then the failure envelope should ultimately change from one with cohesion to one appropriate for a fully granular material such as a dry sand or gravel. Some approaches in the past did not do that, but simply assumed zero strength (as water) when damaged. A scalar damage parameter $D$ describing the accumulation of tensile and/or shear fractures and/or pore crushing from undamaged ($D=0$) to totally damaged ($D=1$) is often used to interpolate between the intact (eq. \ref{eq:Yi}) and the damaged (eq. \ref{eq:Yd}) strength
\begin{equation}
Y=(D-1)Y_i + D Y_d
\label{eq:combined}
\end{equation}
where $Y$ is limited such that $Y\le Y_i$. The damage parameter is computed using a fracture model and it may also be related to porosity models (e.g. \emph{Jutzi et al.} 2008).

Continuum fracture models (e.g. \emph{Grady and Kipp} (1980)) often use an underlying structure with a preexisting Weibull distribution (\emph{Weibull}, 1939) of cracks that grow and coalesce under tensile loading. In these models, cracks grow at a fixed speed $c_g$ once a flaw becomes active. This finite speed of crack grow naturally leads to a rate dependent failure of the material, as it is observed for rocks (see, for example, \emph{Housen and Holsapple}, 1990 for the application to disruptions). In Asteroids III (\emph{Asphaug et al.} 2002), various aspects of the rate and size dependence fracture models were discussed. The implementation of the fracture models in shock physics codes are discussed in e.g. \emph{Benz and Asphaug} (1994, 1995), \emph{Collins et al.} (2004) and \emph{Jutzi et al.} (2008). These models are generally overlaid in addition to the strength envelopes discussed above.

Damage can also increase due to pore crushing during the compression of a porous material. This effect can be taken into account using a (linear) relation between distention $\alpha$ and damage $D$ (e.g. \emph{Jutzi et al.} 2008). 

Damage is usually treated as a scalar parameter. However, others have introduced tensor measures of damage (e.g. \emph{Lubarda and Krajcinovic}, 1993). Due to their complexity those have not yet made their way into impact studies, although tensor damage has been included in some recent codes (e.g. \emph{Owen}, 2010). 

\bigskip
\noindent
\textit{ 4.4.5. Porosity models}
\bigskip

In section 2 we discussed the various scales of porosity and the effects of porosity during an impact, such as the absorption of energy due to compaction, or the reduction of the wave speed. 

Depending on the scale of porosity, it can be modeled as either macroscopic voids (macroporosity) or by using a continuum, sub-resolution material model (microporosity), or as a combination of both as discussed in section 2. Here the size scale for demarcation is a computational cell size. 

A widely used continuum model is the "P-alpha" model (\emph{Herrmann}, 1969; \emph{Carroll and Holt}, 1972), which uses the distention $\alpha$ defined by 
\begin{equation}
\alpha=\rho_s/\rho
\label{eq:distention}
\end{equation}
where $\rho$ is the bulk density of the porous material and $\rho_s$ is the density of the corresponding solid (matrix) material. A crucial assumption in this model is that it is the density and specific internal energy of the matrix material that determines the pressure. That is true when ignoring any energy content related to the porosity such as surface energy. Therefore, the bulk pressure of the porous material $P$ is then related to the pressure in the solid component (matrix) $P_s$:
\begin{equation}
P=\frac{1}{\alpha} P_s(\rho_s,E_s) = \frac{1}{\alpha} P_s(\alpha\rho, E).
\label{meos}
\end{equation}
A significant feature of this model is that any existing EOS for a nonporous material $P_s(\rho_s,E_s)$ can be used as the solid component of a porous material of the same composition. 

A compaction model is required to determine the history of distention as a function of the history of pressure change. It is often assumed that the crushup is independent of shear stress, although \emph{Jutzi et al.} (2008) relate the changing rate of the distention and of the deviatoric stress tensor. The form of the model for the historic $P$-$\alpha$ Herrmann form includes a crush curve $\alpha=\alpha_c(P)$, at which pressure increases will decrease the distention, defining the crush-up of the material. In addition, the porous model defines unloading elastic curves. Those apply until the stress state again reaches a failure boundary.  \emph{Herrmann} (1969) does not discuss that feature. The curves for loading and unloading are obtained by pressure tests that load and unload the material and are defined in the model with appropriate algebraic forms. The quadratic form assumed by Herrmann is not well suited to geological materials, but is easily changed. For instance, \emph{Jutzi et al.} (2008, 2009) use a combination of two power-law functions to successfully reproduce the crush curve of pumice (Fig. 1 in \emph{Jutzi et al.}, 2009).  

Since the EOS has the underlying density as a function of both the pressure and the internal energy,  the crush curve should include that also, so \emph{Holsapple} (2008) suggests that $\alpha=\alpha_c(P,E)$. In particular, the crush strength should become zero when the thermodynamic state is at melt. That feature is not yet common in models.

Elastic unloading only occurs until the stress state reaches some other point on the enclosing failure envelope.  When shear stress is present, that can occur at positive pressure, but it will be at pure tensile pressure if the shear is zero and cohesion is present. At those states, increasing porosity will occur: the occurrence with shear is the physical process of dilation in shear flow.  Such an approach has been implemented recently in the iSALE code (\emph{Collins} 2014).

The Herrmann approach to solving the history of distention following these paths used a time sub-cycling which was essentially a forward differencing.  The difficulties in that numerical method motivated an alternative model which determines the distention as a function of volumetric strain, it was presented by \emph{W\"unnemann et al.} (2006). The so-called "epsilon-alpha" model addresses the above mentioned problem of the iterations in the "P-alpha" model due to the interdependency of pressure $P$ and distention $\alpha$. When implementing eq. \ref{meos} and the crushing curve $\alpha=\alpha(P)$ into a hydrocode  for a given time step $t$, $\alpha_{t+1}$ must be known to derive $P_{t+1}$, but $\alpha_{t+1}=f(P_{t+1})$. A common solution is to use small sub-cycles to iterate the new $P(t+1)$ value. This method requires extra computation time and may be numerically unstable under certain circumstances. The problem is solved in the "epsilon-alpha"-model using the volumetric strain $\epsilon_V$ to determine the crushing of pore space: $\alpha=\alpha(\epsilon_V)$. The "epsilon-alpha" model distinguishes four compaction regimes, where the rate of compaction $d\alpha/d\epsilon_V$ is calculated according to elastic compaction ($\epsilon_V<\epsilon_e$), exponential compaction ($\epsilon_e<\epsilon_V<\epsilon_x$), power-law compaction ($\epsilon_x<\epsilon_V<\epsilon_c$), and the fully consolidated state ($\epsilon_V>\epsilon_c$).

The improved model (\emph{Collins et al.}, 2011) removed two shortcomings of the initial model.  First, it accounts for the fact that the speed of sound of the pristine porous material can be substantially lower than the elastic wave speed in the solid material. A simple linear relationship for $c(\alpha)$ interpolating between the speed of sound of the fully compacted material and the initial porous material is assumed.
Secondly, the improved version distinguishes between thermal and mechanical strains. The improvement uses an approximation about the equation of state, so further calls to the EOS subroutine are not needed. In contrast to the original "epsilon-alpha" model the improved version is also applicable for highly porous material where material is heated extremely due to the compaction of pore space (PdV-work) resulting in thermal expansion of the solid component.

An alternative idea that retains the $P$-$\alpha$  form was outlined in \emph{Holsapple} (2008). It is based on a Newton-Raphson approach rather than the forward differencing and eliminates the numerical problems in the original formulation.

\bigskip
\noindent
\textit{ 4.4.6. Fluidization models}
\bigskip

In the opinion of some researchers, the constitutive models describing the strengths of rocks cannot explain the formation of complex crater structures with central peaks, where originally deep-seated material was uplifted several kilometers. And they do not explain the shallow outer wall slopes that are well below the angle of repose of the models. It is assumed that an almost fluid-like rheology of matter during crater formation is required, and it is from some weakening mechanism that lasts only temporarily. The latter is an important constraint as steep, almost vertically standing flanks of central peaks can only be explained if rocks almost return to their initial strength. 

A successful approach to solve the problem of a temporary fluid-like rheology of rocks during crater formation was suggested by \emph{Melosh} (1979). He proposed that heavily fractured and brecciated rocks behave like a granular flow excited by acoustic vibrations that have been interpreted to be observed in the ground motions generated by nuclear explosions. The fluid-like nonlinear rheology decays as the amplitude of the vibration attenuates. The model requires that the wave-length of the acoustic signal is comparable to the size of the fragments. The original acoustic fluidization model has been simplified in the so-called block model where the acoustically fluidized material behavior is described by a Bingham model (\emph{Melosh and Ivanov}, 1999). The Bingham viscosity is proportional to the block size and the Bingham cohesion depends on the amplitude of the vibration. The latter is a function of time as the amplitude of the acoustic wave decays. A shortcoming of the acoustic fluidization or block model is the fact that neither block size nor decay time are know and can only be estimated for the size of a given structure. \emph{W\"unnemann and Ivanov} (2003) suggested a heuristic linear relationship for both block size and decay time with the size of the impactor. They assume that larger impactors produce larger blocks and longer waves that attenuate slower than in case of smaller impacts where material is fractured into smaller fragments and the acoustic signal  vanishes much quicker. The linear scaling of the block model parameters has been calibrated against the observed depth diameter ratios of the crater record of the moon (\emph{W\"unnemann and Ivanov}, 2003) and on satellites (\emph{Bray et al.}, 2014). 
   
\begin{figure}[t!]
 \epsscale{0.9}
 \plotone{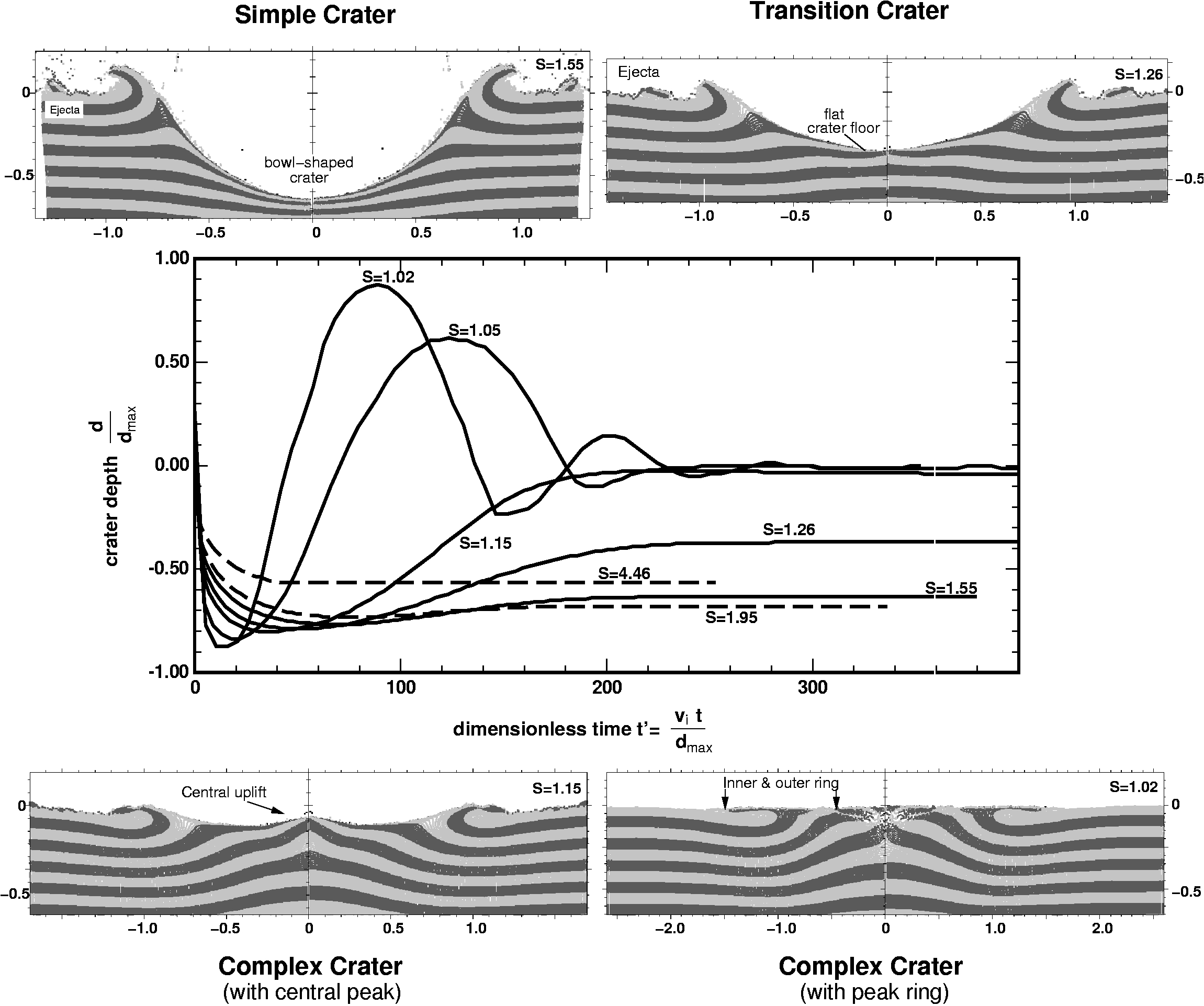}
 \caption{\small Crater depth versus time for different impact conditions (S = 1.02-4.46). The length parameter is normalized by the maximum crater depth $d_{max}$, time is scaled by the ratio of impact velocity $v_i$ and $d_{max}$. The no-fluidized rheology is given by a Drucker-Prager yield strength envelope where $Y=Y_0+\phi p$, with $Y_0$=25 GPa and $\phi$=0.1. Examples of the resulting crater morphology for different $S$-values (different projectile sizes) are shown above and below the diagram. Note, all length scales are normalized by $d_{max}$ (taken from \emph{W\"unnemann and Ivanov}, 2003).}  
\label{figure3}
 \end{figure}

Figure \ref{figure3} shows a simulation of the gradual transition with increasing crater size from simple to complex crater morphology, using that model. The parameter $S$ is given by the ratio between strength $Y$ and the hydrostatic pressure at maximum crater depth $d_{max}$: $S=Y/(\rho g d_{max)}$. The diagram illustrates the increasing crater depth while the transient cavity is growing and the decrease of crater depth when the crater floor collapses and starts to rise forming a complex crater with a central peak. According to some simple estimates by \emph{Melosh} (1979) the critical value for crater collapse is given by $S=0.25$ which corresponds to strength of the rocks of a few MPa, approx. one order of magnitude smaller than typical values for rocks. The application of the block model allows to assume realistic strength values for crater collapse at the observed transition diameter between simple and complex crater morphology on different planets. 

The approach described above is well established and was successfully applied in the past in many studies to model the formation of complex craters. However, there is an ongoing debate about the actual underlying physical mechanism that causes the temporary weakening. 
Holsapple (2004a, 2013) suggested that existing strength theories, in their complete form, can model the late time readjustments discussed above, but his approach has not yet been fully vetted.  Some important evidence for the related problem of the collapse and runout of a granular cliff is given in section 4.6.3 below.  

\bigskip
\noindent
\textbf{ 4.5 Examples of hydrocodes}
\bigskip

Various approaches are used in shock physics codes to solve the continuum system of partial differential equations detailed in section 4.3. The equations may be cast in a ``Lagrangian" reference frame that follows the material or in an ``Eulerian" reference frame that is fixed in space. The continuum codes include finite difference, finite element, and SPH methods. Each numerical method has its own strengths and weaknesses. In a recent review by \emph{Collins et al.} (2013), the various methods are presented and numerical issues like resolution, the treatment of shocks, multi-material approaches, etc. are discussed. Here we present a few examples of commonly used hydrocodes. For each method, the specific material models used and the pros and cons are indicated.

\bigskip
\noindent
\textit{ 4.5.1. Grid based codes}
\bigskip

Most numerical simulations of planetary collision processes use Eulerian grids, because large deformations are difficult to track in the Lagrangian approach.

They  typically use a two-step approach, where the first step is the Lagrangian step where the deformation of the grid according to a given velocity field is calculated, and then there is a second step that maps the grid back on its original location in space. The remapping requires special treatment of material boundaries which is usually solved by tracking or reconstructing the interface between different matter (for a discussion of different methods of interface tracking/reconstruction see \emph{Elbeshausen and W\"unnemann}, 2010). The main advantages of grid-based codes are:
\begin{itemize}
\item Lagrangian, Eulerian or an arbitrary spatiotemporal transition between both reference frames is possible
\item High resolution that may be increased locally by the adaptive mesh refinement (AMR) method (\emph{Berger and Oliger}, 1984; \emph{Berger and Colella}, 1989) employed in CTH
\item Adequate treatment of solid state deformation, liquid flow, and gas expansion including appropriate rheology models
\item Coupling of the grid-based methods with massless Lagrangian tracer particles to track the spatiotemporal thermodynamic history of matter
\item Relatively straight forward implementation of a variety of Equation of State (EOS) models 
\end{itemize}

Grid-based models come also with some disadvantages:
\begin{itemize}
\item In Eulerian mode, material interfaces are not tracked perfectly and bulk properties must be defined for mixed cells
\item Matter is treated as continuum which makes it difficult to treat fragmentation properly. Separate fragments tend to be underresolved unless AMR is employed
\item Although multi-material handling is possible, mostly the codes do not deal with multi-phase processes where each phase moves at its own speed resulting in mixing of matter. Such processes are in particular important for the interaction of solid state fragments with an expanding vapor plume. 
\end{itemize}

In this paper we mention three shock physics codes widely used in planetary science. A much more complete overview of available hydrocodes is given in \emph{Pierazzo et al.}, (2008).

The most advanced software package among those is the Sandia National Laboratories CTH code (\emph{McGlaun et al.}, 1990); However, because of its military uses, this code is limited in its availability to US citizens. Besides highly advanced material models it also includes AMR and self-gravity. For disruptive large scale collision processes such as the moon-forming impact event self-gravity is essential (\emph{Canup et al.}, 2013). \emph{Canup et al.} (2013) also compare results produced by different approaches such as grid-based models (CTH) and mesh-free codes (SPH, see below). \\
Other widely used hydrocodes are SOVA (\emph{Shuvalov},1999) and iSALE (\emph{Amsden et al.}, 1980; \emph{Collins et al.}, 2004; \emph{W\"unnemann et al.}, 2006; \emph{Elbeshausen et al.}, 2009) that are, with some limitations, freely available for scientific purpose. SOVA  contains a specific routine to deal with the interaction of lithic and molten ejecta and the vapor plume. As mixing of all phases occurs it may be better called ejecta plume. This process requires so-called multi-phase hydrodynamics where at least two different phases (ejecta and vapor) travel at different speeds interacting and mixing with each other. SOVA addresses this problem by introducing representative tracer particles where each tracer represents a certain number of fragments of a given size. The tracers exchange momentum and energy with the surrounding gas, but not with each other. The approach only works if the volume of fragments is small relative to the volume of the gas. The method has been successfully applied to model dusty flows (\emph{Shuvalov}, 1999), tektite formation and deposition (\emph{St\"offler et al.}, 2002), and the development of Chicxulub distal ejecta (\emph{Artemieva and Morgan}, 2009). \\
The iSALE hydrocode is based on the SALE hydrocode solution algorithm (\emph{Amsden et al.}, 1980). To simulate hypervelocity impact processes in solid materials SALE was modified to include an elasto-plastic constitutive model, fragmentation models, various EoS (Tillotson and ANEOS), and multiple materials (\emph{Melosh et al.}, 1992; \emph{Ivanov et al.}, 1997). More recent improvements include a modified strength model (\emph{Collins et al.}, 2004) and a porosity compaction model (\emph{W\"unnemann et al.}, 2006; \emph{Collins et al.}, 2011). The 3D version uses a numerical solver as described in \emph{Hirt et al.} (1974). The development history of iSALE-3D is described in \emph{Elbeshausen et al.} (2009). iSALE has been used to model the collision of highly porous planetesimals (\emph{Davison et al.}, 2012), impacts on the surface of Vesta (\emph{Krohn et al.}, 2014) and the formation of large craters on Lutetia (\emph{Cremonese et al.}, 2012), and the effect of an oblique impact angle on crater formation (\emph{Elbeshausen et al.}, 2013; \emph{Collins et al.}, 2012; \emph{Elbeshausen et al.}, 2009).

\bigskip
\noindent
\textit{ 4.5.2. Smoothed Particle Hydrodynamics (SPH) codes}
\bigskip

Smoothed particle hydrodynamics (SPH) is a meshless  continuum Lagrangian approach (and in spite of its name is not a "particle" code). In the basic SPH method, approximate numerical solutions of the fluid dynamics equations are obtained by replacing  the fluid with a set of particles (see e.g., \emph{Gingold and Monaghan }1977, \emph{Monaghan} 2012). The properties of these particles are smoothed over a certain length $h$ by a kernel function. That provides the link to a continuum approach. As the flow evolves, material mass moves with the particle and local density is calculated based on the proximity of nearby particles.
The main advantages of this mesh-free method are:

\begin{itemize}
  \item SPH is a very robust scheme. It is 'straightforward'  to add new physics. 
 \item Large deformations of the simulated objects are handled easily.
 \item The description of free surfaces is trivial.
 \item Due to the particle nature of the method, fragmentation is modeled naturally by the separation of particle clusters.
\end{itemize}

Disadvantages of the SPH method include:
\begin{itemize}
 \item The spatial resolution is typically lower than in grid-based simulations.
 \item Boundary conditions and discontinuities (e.g. material interfaces) are complicated to handle.
 \item It is mostly applied in 3D, which often requires computationally expensive million-particle calculations 
\end{itemize}

In the past years, the range of applications of the SPH algorithm has increased significantly. A number of SPH codes have been recently developed to study problems in planetary sciences. Examples are a SPH code for the modeling of pre-planetesimals (\emph{Geretshauser et al.} 2011), SPHERAL (\emph{Shapiro, et al.} 1996, \emph{Owen} 2010), GADGET (\emph{Springel}, 2005; \emph{Marcus et al.} 2009), SPHLATCH (\emph{Reufer et al.}, 2012) or a GPU-based SPH code (\emph{Kaplinger et al.} 2013).

A widely used SPH code to model collisions among rocky bodies was developed by {\em Benz and Asphaug} (1994, 1995). This code was further extended by {\em Jutzi et al.} (2008, 2013) with the goal to realistically model rocky bodies with various internal structures (see section 4.5 for a comparison to laboratory experiments). The most recent version of this code (see \emph{Jutzi}, 2014) includes a pressure depended strength model as outlined in section 4.4.2, a tensile fracture model (section 4.4.3) and a porosity model based on the P-alpha model (section 4.4.5) and self-gravity. Friction is modeled either using the Coulomb dry friction law (eq. \ref{eq:Yd}) or the rate-dependent model suggested by \emph{Jop et al.} (2006). Self-gravity (section 4.3) is included as well. Recent applications of this code in asteroid studies are presented in section 5 and in the chapter by \emph{Asphaug et al.} (this volume).

\bigskip
\noindent
\textbf{ 4.6 Comparison to laboratory experiments}
\bigskip

An important step in the development of numerical methods (which includes the implementation of complex material models) is the validation against laboratory experiments. As rigorous testing is limited by the availability of appropriate experiments, including measurements during the highly dynamic processes, a preliminary step is benchmarking of different codes against each other. An example for a very successful benchmark and validation study is given in \emph{Pierazzo et al.} (2008) where the modelling results of specific test problems from 8 different hydrocodes were compared against each other and validated against laboratory cratering experiments, but only for  water and aluminium.  

Here we provide some additional examples dealing with the fracturing and fragmentation resulting from collisional processes, and a granular flow problem.

\bigskip
\noindent
\textit{ 4.6.1 Modeling the fragmentation of porous pumice}
\bigskip

\begin{figure}[t!]
 \epsscale{0.9}
 \plotone{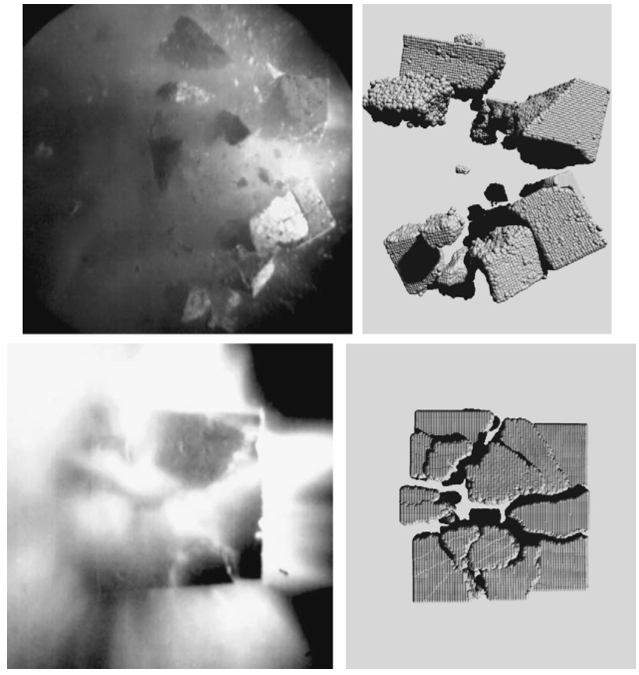}
 \caption{\small Comparison between laboratory impact experiments (left) and SPH code calculations (right). The targets are pumice cubes of size of 7 cm with a porosity of $\sim$ 70\%. The impact experiments were performed at the Institute of Space and Astronautical Science (ISAS) of the Japan Aerospace Exploration Agency (JAXA) using a two-stage light-gas gun. Two impacts are shown at different times: t = 8 ms (top) and t = 1.5 ms (bottom). From Jutzi et al. (2009)}  
\label{figure4}
 \end{figure}

Figure \ref{figure4} shows a comparison of an SPH code calculation (using the SPH method as described in section 4.5.2) to laboratory impact experiments. The target used in the experiments is porous pumice with a porosity of $\sim$ 70\%. The fragment size distributions resulting from the impacts with velocities ranging from 2 to 4 km/s could be well reproduced in the SPH code calculations (see Jutzi et al. 2009 for details). This example illustrates the capabilities of the SPH method to model impact fragmentation. 

 \bigskip
\noindent
\textit{ 4.6.2 Fracturing in cratering experiments}
\bigskip

\begin{figure}[t!]
 \epsscale{0.9}
 \plotone{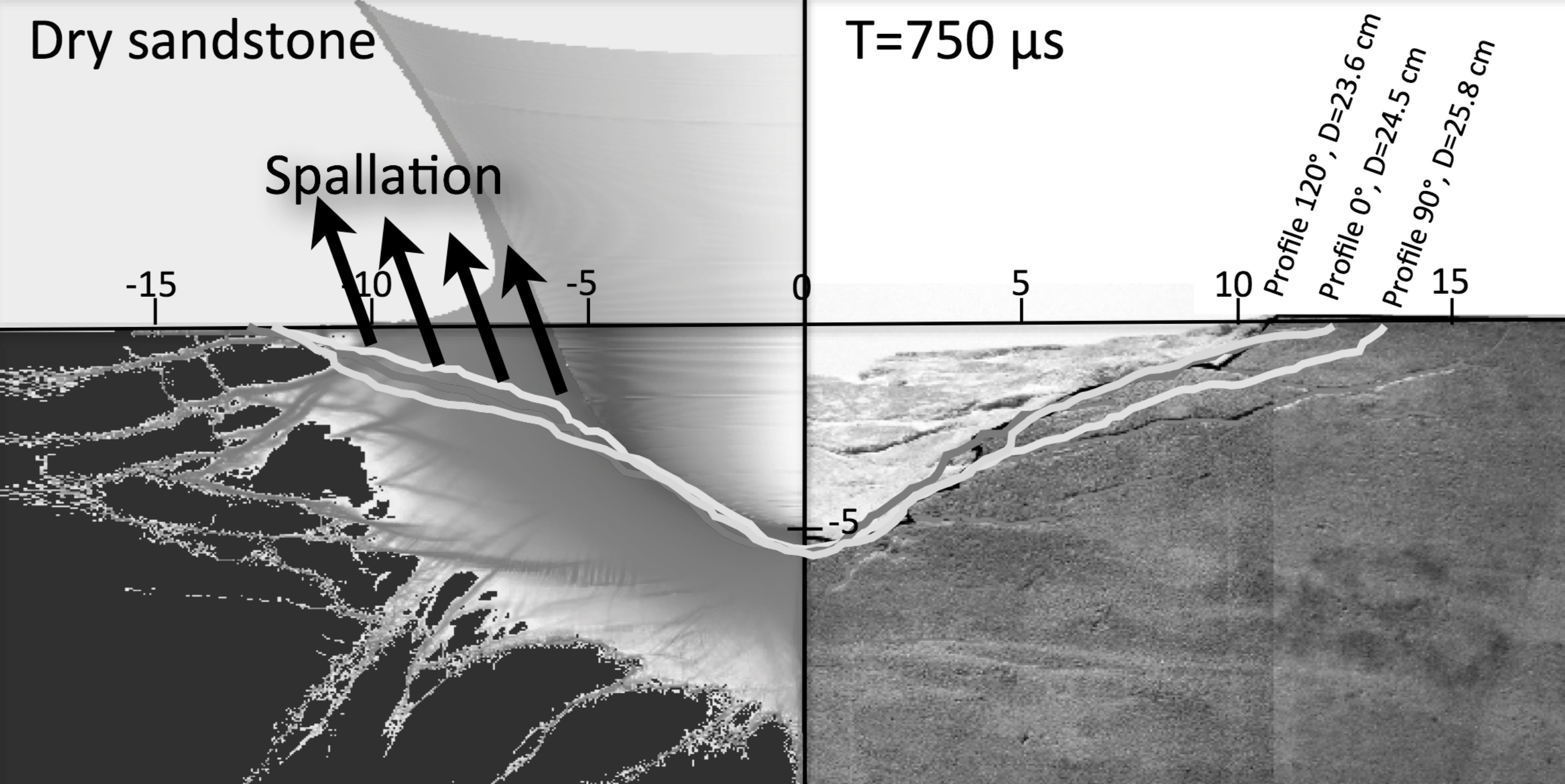}
 \caption{\small Comparison between laboratory cratering experiments (right) and iSALE simulation (left). The target was a 100x100x50cm sandstone block with 20\% porosity. The so-called Seeberger sandstone is composed of 97 w\% SiO2 with an unconfined compressive strength of $\sim$ 60 MPa. The projectile was a 1cm iron sphere impacting at 5.4 km/s (for further details see \emph{Kenkmann et al.} (2011). Right: The thin gray lines indicate macroscopic flaws, the thick gray lines indicate crater profiles in different directions. Left: snapshot of the numerical model showing contours of the damage parameter. The black lines indicate the zone where tensile failure occurs and material spalls off the surface. Courtesy by the MEMIN-team}  
\label{figure5}
 \end{figure}

Fracturing caused by hypervelocity impact does not necessarily result in a complete disruption of the target. Figure \ref{figure5} shows the cross-section of the crater  that was formed by the 5.4 km/s impact of a 1cm diameter iron projectile in $\sim$20\% porous sandstone. The experiment was carried out in the framework of the so-called MEMIN (Multidisciplinary Experimental and Modeling Impact Research Network) project aiming at the validation of hydrocode modelling of hypervelocity impact processes (see Kenkmann et al., 2013 and papers therein). The left frame of Figure \ref{figure5} shows a snapshot of the iSALE cratering model at 750 $\mu s$. The model includes the $\epsilon-\alpha$ porosity compaction model and a strength and damage model as described in (\emph{Collins et al.}, 2004). Some static strength parameters of the sandstone were available (\emph{Kenkmann et al.}, 2011), others such as crushing strength of the porous sandstone were estimated or adjusted to match the observed crater depth. 

A more detailed calibration and validation of the crushing of pore space and sandstone is presented in (\emph{G\"uldemeister et al.}, 2013; \emph{Kowitz et al.}, 2013). The crater in the experiment is enlarged by spallation. Modelling of the actual spall of material from the surface is not included in the iSALE model; however, the spall zone is indicated by cells that have experienced failure.

The damage zone in the model (red and orange contours represent the damage parameter as introduced above) underneath the crater shows some qualitative similarities with the flaws highlighted in red on the cross-section of the cratering experiment. On the scale of individual grains it was shown that the zone where pore space was crushed or fracturing occurs in grains extends approximately to the same distance as in the model.     

  \bigskip
\noindent
\textit{ 4.6.3 Cliff collapse problem}
\bigskip

 The cliff collapse problem is a useful test case for the pressure-dependent models used in code calculations, at least for low pressure. In that problem, an initial vertical "cliff" of material is suddenly released, and the mass falls and flows out to considerable lengths.

This problem is of special interest for cratering because the same requirements of a temporary fluidization mechanisms have been asserted as necessary for the mechanics of landslides. The final angles of the runouts are typically $<10^\circ$,  far below the angle of repose of geological materials.

\emph{Holsapple} (2013) presented detailed theoretical arguments and numerical calculations of that landslide problem, using only the standard Drucker-Prager soil model with a 35$^\circ$ angle of friction.  He concludes that no additional ad-hoc models are required to reproduce observed laboratory results. Specifically, his numerical simulations correctly reproduce laboratory experiments with final slopes  $<10^\circ$.  A detailed investigation of those simulations explains the reason.  It lies in the fact that the "slopes" allowed by the angle of repose argument are not violated, it is just that during the highly dynamic flows, the "slopes" must be measured relative to the combination of the local gravity direction and the direction of the inertial forces.

\emph{Jutzi} (2014) also presented SPH calculations of that problem. He compared two models, one with a constant coefficient of friction (eq. \ref{eq:Yd}) and one using a rate-dependent and a particle size dependent relation defined in an "inertial number" as suggested by \emph{Jop et al.} (2006). It was found that both models reproduce the experiments very well. The two models lead to the same results because in this granular flow regime, the inertial number stays small and therefore the rate dependency is negligible. This finding provides further evidence that the global outcome of such events is well reproduced by using a simple Coulomb dry friction law with a single parameter $\mu_d$, although at higher pressures a non-linear model is undoubtedly necessary.

These examples show that the cliff collapse problem and the resulting runouts (at laboratory conditions) can be well reproduced with continuum codes using conventional rock mechanics. However, there is evidence that additional weakening mechanism (section 4.4.6) are necessary to explain other cases, such as the very long runout slides observed on Mars (e.g. \emph{Lucas and Mangeney}, 2007; \emph{Harrison and Grimm}, 2003).

\bigskip 
\centerline{\textbf{ 5. APPLICATIONS}}
\bigskip 

In this section, some examples of recent modeling are presented which illustrate the various the effects of the material properties, target structures and impact conditions on the outcome of a collision.  

\bigskip
\noindent
\textbf{ 5.1. Scaling laws from numerical modeling}
\bigskip

\bigskip
\noindent
\textit{ 5.1.1. Cratering regime}
\bigskip

\begin{figure}[t]
 \epsscale{0.9}
 \plotone{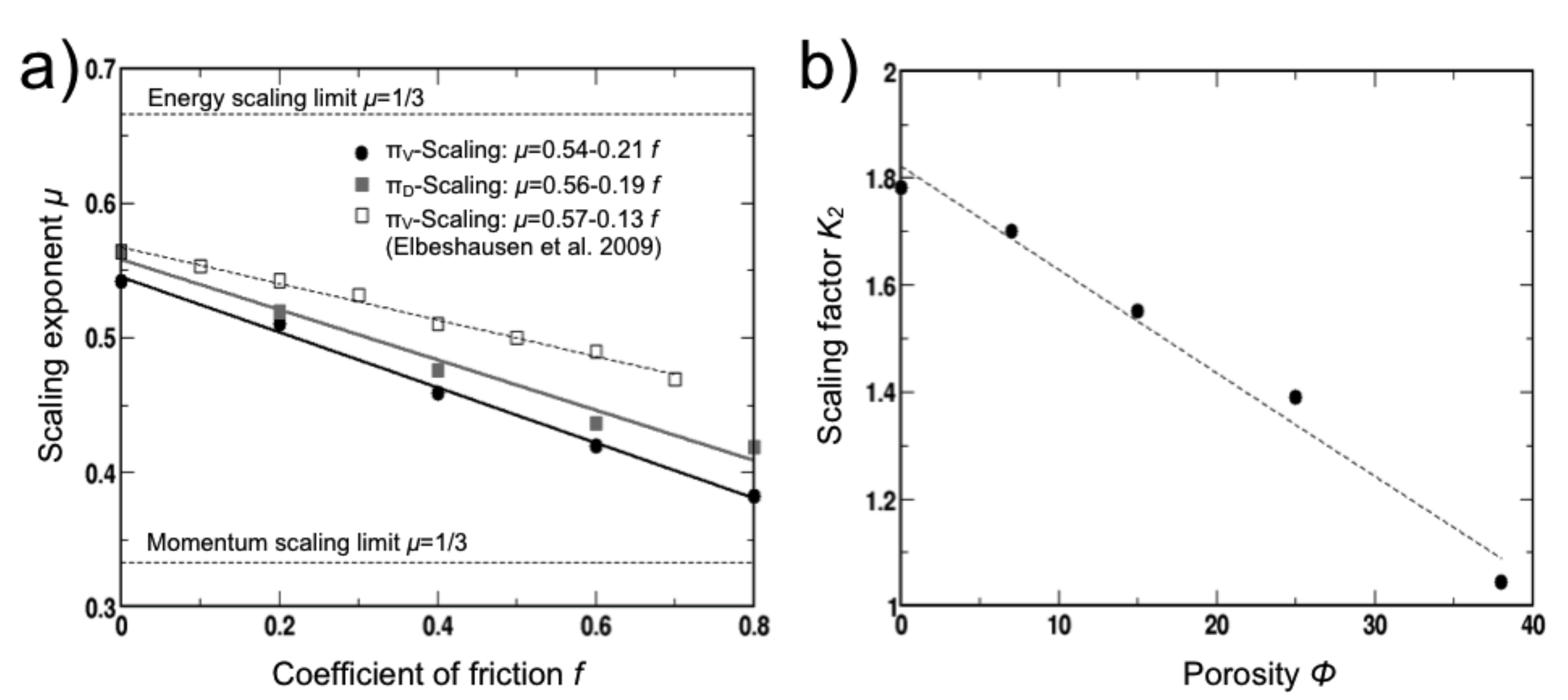}
 \caption{\small Scaling parameters $\mu$ and $K_2$ as a function of coefficient of friction $f$ and porosity $\phi$. (a) The scaling exponent $\mu$ was determined by numerical impact experiments into material with different coefficient of friction, zero cohesion, and zero porosity. The dashed line was taken from \emph{Elbeshausen et al.} (2009). (b) The scaling factor $K_2$ was determined by numerical impact experiments into material with different porosities, a coefficient of friction $f=0.8$, and zero cohesion.} \label{figure6}
 \end{figure}

\begin{figure}[t!]
 \epsscale{0.9}
 \plotone{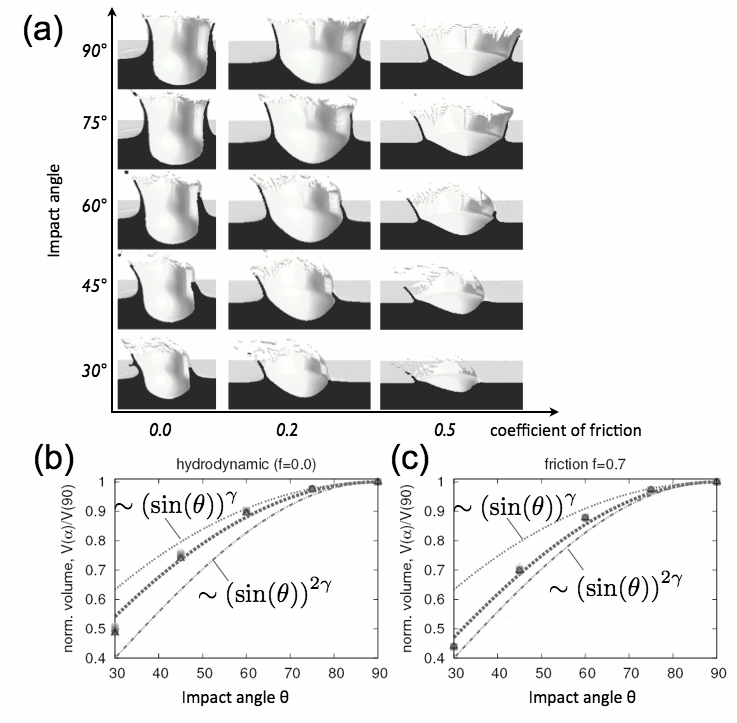}
 \caption{\small Snapshots of the approximate transient crater for different impact angle and coefficient of friction. Relative change of the crater volume $V_{\theta}$ normalized by crater volume for an equivalent 90 degree impact $V_{90}$  with  impact angle for (b) $f = 0$ (fluid-like target rheology) and (c) $f = 0.7$ (sand-like target rheology). Note, the results from numerical models over a range of different $\pi_2$ values fall in between two lines calculated according to the vertical velocity component approximation ($\sim \sin(\theta)^{2\gamma}$; lower curve),  and another line ($\sim \sin(\theta)^\gamma$). Taken from  \emph{Elbeshausen et al.}, (2009). } \label{figure7}
 \end{figure}

Simple power-law relationships, so-called scaling laws, have been proposed to relate the size of the transient crater with the impact parameters such as the mass and velocity of the impactor (see section 3). Those have been shown to be a theoretical consequence of the point source assumption, as mentioned above. The scaling parameters in these equations (the so-called velocity exponent $\mu$ and a proportionality factor that may be called $K$) have been determined by laboratory experiments, mainly in sand or other granular materials. One way to determine the scaling parameters is to vary the so-called gravity-scaled size of an impact event $\pi_2=(L g)/v^2$, with the diameter $L$ of the impactor, or the gravity $g$, or the impact velocity $v$, over a large as possible range. For example, one can increase the gravity using a geotechnic centrifuge, and decrease the gravity using drop towers. The geotechnic centrifuge method was developed both for explosive cratering and for hypervelocity impact cratering by \emph{Schmidt} (1977) and by \emph{Schmidt and Holsapple} (1980). Since gravity can be varied over two orders of magnitude, a 500 G centrifuge test models 1G events with a 500 times larger scale.  A meter-sized test at 500G has the same physics as a 0.5 km event at 1 G. However,  there is no affect of the increased gravity at those magnitudes except for materials with very little cohesion, so studies are primarily limited to granular materials such as sand. Although the properties of the target material such as porosity or internal friction are somewhat under the experimenter's control, varying these parameters independently is difficult.

An alternative approach is to use numerical simulations to carry out numerical experiments which are not limited by experimental constraints in the laboratory. Of course, due to the numerous modeling questions  mentioned above, those are not without their own uncertainties.  And, as long as the point-source approximation is valid, they cannot contradict the theoretical scaling forms, but they can identify the unknown scaling coefficients.  And they allow extensions of the parameter space from laboratory scale to natural dimensions under any given gravity regimes of planets, moons, and asteroids. The effect of material properties of the target such as porosity, friction, cohesion can be investigated independently (\emph{W\"unnemann et al.}, 2010) and also the effect of the angle of incidence can be investigated (\emph{Elbeshausen et al.}, 2009). In these studies it was shown that the velocity exponent $\mu$ depends on the coefficient of friction $f$ of the target material (Figure \ref{figure6}) and the scaling factor $K$ may be expressed as a function of porosity $\phi$.

To incorporate the angle  of impact $\theta$ into the scaling equations it was suggested by \emph{Chapman and McKinnon} (1986) and others based on impact experiments in sand to replace the impact velocity $v_{\theta}$ by the vertical component of the impact velocity $v_{\perp}$.  In this case, for instance crater volume scales with the sine of the impact angle raised to the power $2\gamma$ where $\gamma=3\mu/(2 + \mu)$. Extensive parameter studies using models of impacts at different angles into granular targets with varying coefficient of friction (Figure \ref{figure7}a) show that only for a typical coefficient of friction of $f=0.7$ this approximation holds true (Figure \ref{figure7}c, \emph{Elbeshausen et al.}, 2009). The vertical velocity component approximation also provides good estimates of crater volume and diameter for strength dominated craters in cohesive ductile targets such as metals (\emph{Davison et al., 2011}). For impacts in granular targets with smaller coefficient of friction than typical for sand the size of the resulting crater is underestimated by this simple assumption (Figure \ref{figure7}b).  Details of the impact may not be so easily matched using this assumption.

\bigskip
\noindent
\textit{ 5.1.2. Disruption regime}
\bigskip

To characterize the outcome of a disruptive collision, the critical specific impact energy $Q^*_D$ which results in the escape of half of the target's mass in a collision is often used. The parameter $Q^*_D$ is called  the catastrophic impact energy threshold (also called the dispersion threshold). The specific impact energy is often defined as $Q = 0.5 m_p v_p^2 / M_T$, where $m_p$, $v_p$ and $M_T$ are the mass and speed of the projectile and the mass of the target, respectively. The catastrophic disruption threshold $Q^*_D$ is then given by the specific impact energy leading to a largest (reaccumulated) fragment $M_{lr}$ containing 50\% of the original target's mass. In recent studies (e.g. \emph{Stewart and Leinhardt, 2009; Leinhardt and Stewart 2012}), a more general definition of the specific impact energy was proposed which also takes the mass of the impactor into account, which can be substantial in very low velocity impacts of near-equal-sized bodies.  The corresponding radius $R_{C1}$ is then defined as the spherical radius of the combined projectile and target masses at a density of 1 g cm$^{-3}$. According to this new definition, the catastrophic disruption threshold is then called $Q^*_{RD}$. 

Values of $Q^*_D$ (or $Q^*_{RD}$) have been estimated using both laboratory and numerical hydrocode experiments (see e.g. \emph{Holsapple et al., 2002; Asphaug et al. 2002}).
For high velocity asteroid collisions, the first suite of numerical calculations aimed at characterizing the catastrophic disruption threshold in both the strength regime and the gravity regime was performed by \emph{Benz and Asphaug} (1999), who used a smoothed particle hydrodynamics (SPH) code \emph{Benz and Asphaug} (1994,1995) to simulate the breakup of basalt and icy bodies from centimeters-scale to hundreds kilometers in diameter. More recently, \emph{Leinhardt and Stewart} (2009)  computed $Q^*_D$ curves using the hydro code CTH \emph{McGlaun, 1990} to compute the fragmentation phase and the N-body code \emph{pkdgrav} to compute the subsequent gravitational evolution of the fragments. In this study, the dependency of $Q^*_D$ on the strength of the target was investigated. In a recent study by \emph{Jutzi et al.} (2010),  the effect of target porosity on $Q^*_D$ was investigated using an extended version of the SPH code (\emph{Jutzi et al.}, 2008).  In this study, the size and velocity distribution of the fragments was computed as well, using the \emph{pkdgrav} code. \emph{Benavidez et al.} (2012) performed a study of a large number of  collisions among $R_t$ = 50 km rubble pile bodies using the original SPH code by  \emph{Benz and Asphaug} (1994,1995). 
As discussed in section 3, \emph{Leinhardt and Stewart} (2012) proposed general scaling laws for collisions among gravity dominated bodies. 

In the chapter by \emph{Asphaug et al.} (this volume), a recent systematic study of the relative effects of various asteroid properties on the disruption threshold $Q^*_D$ (Jutzi, 2014) is presented.

\bigskip
\noindent
\textit{ 5.2. Momentum transfer in small impacts}
\bigskip

The study of the momentum transferred in an specific impact on an asteroid as a function of impact conditions and the internal structure  is crucial for performance assessment of the kinetic impactor concept of deflecting a potentially hazardous asteroid from its trajectory \emph{(see \emph{Harris et al.} this volume)}. The momentum transfer is characterized by the so-called momentum multiplication factor $\beta$, which has been introduced to define the momentum imparted to an asteroid in terms of the momentum of the impactor: 
\begin{equation}
\beta= 1 +p_{ej}/(M_p v_p),
\end{equation}\label{eq:beta}
where $p_{ej}$ is the escaping ejecta momentum, and $M_p$ and $v_p$ are the mass and velocity of the impactor, respectively (see e.g. \emph{Holsapple}, 2004b).
In the limiting case of an impact which produces no escaping ejecta, $\beta= 1$, and the momentum transferred corresponds to the momentum of the projectile (inelastic collision). However, in the case of an impact which produces a lot of material (ejected in the opposite direction) with velocities  larger than the escape velocity, we can have $\beta \gg 1$ due to the contribution of $p_{ej}$. 
Using the $\beta$ factor, the resulting momentum of the target $\Delta P_t$ (along the impact direction) can be obtained by 
\begin{equation}
\Delta P_t = \beta P_p,
\end{equation}
where $P_p = m_p v_p$ is the momentum of the projectile.

It is important to note that only fragments with ejection velocities larger than the escape velocity of the target asteroid, $v_{eject} > v_{esc}$, escape the body and contribute to the momentum transfer. Moreover, the ejected material is slowed down during its escape and its trajectory is changed due to the gravitational attraction of the target. Therefore, to compute the escaping ejecta momentum, the value and angle of the velocity at infinity have to be used (see e.g. \emph{Holsapple and Housen,} 2012; \emph{Cheng}, 2012; \emph{Jutzi and Michel}, 2014). 

\emph{Holsapple and Housen} (2012)  presented scaling laws to extrapolate measurements of the momentum transfer in laboratory experiments to asteroid scales.

Recent code calculations of the momentum transfer efficiency in impacts on asteroids for conditions typical for a kinetic impactor were performed by \emph{Holsapple and Housen} (2013a) and \emph{Jutzi and Michel} (2014). \emph{Holsapple and Housen} (2013a) presented numerical simulations both for porous and non-porous materials, and obtained very good agreement with their laboratory results.

 In the study by \emph{Jutzi and Michel} (2014), the effect of different degrees and scales of porosity on the momentum multiplication factor $\beta$ was investigated for a range of impact conditions.

Two kinds of porous target structures were considered by  \emph{Jutzi and Michel} (2014): (a)  homogeneous: microporous only, and (b) heterogeneous: both micro- and macro-porous. For both structures, the microporous part of the material has a porosity of 50\%. For structure (b), in addition to the microporosity, macroscopic cracks with a size scale of $\sim~$0.3 m are randomly distributed in the target. The resulting total macroscopic void fraction  is 10 $\%$. 

The impact simulations were performed using a Smooth Particle Hydrodynamics (SPH) impact code (e.g. \emph{Benz and Asphaug} 1994,1995; \emph{Jutzi et al.} 2008, \emph{Jutzi}, 2014) as presented in section 4.5.2.

\begin{figure}[t]
 \epsscale{0.9}
 \plotone{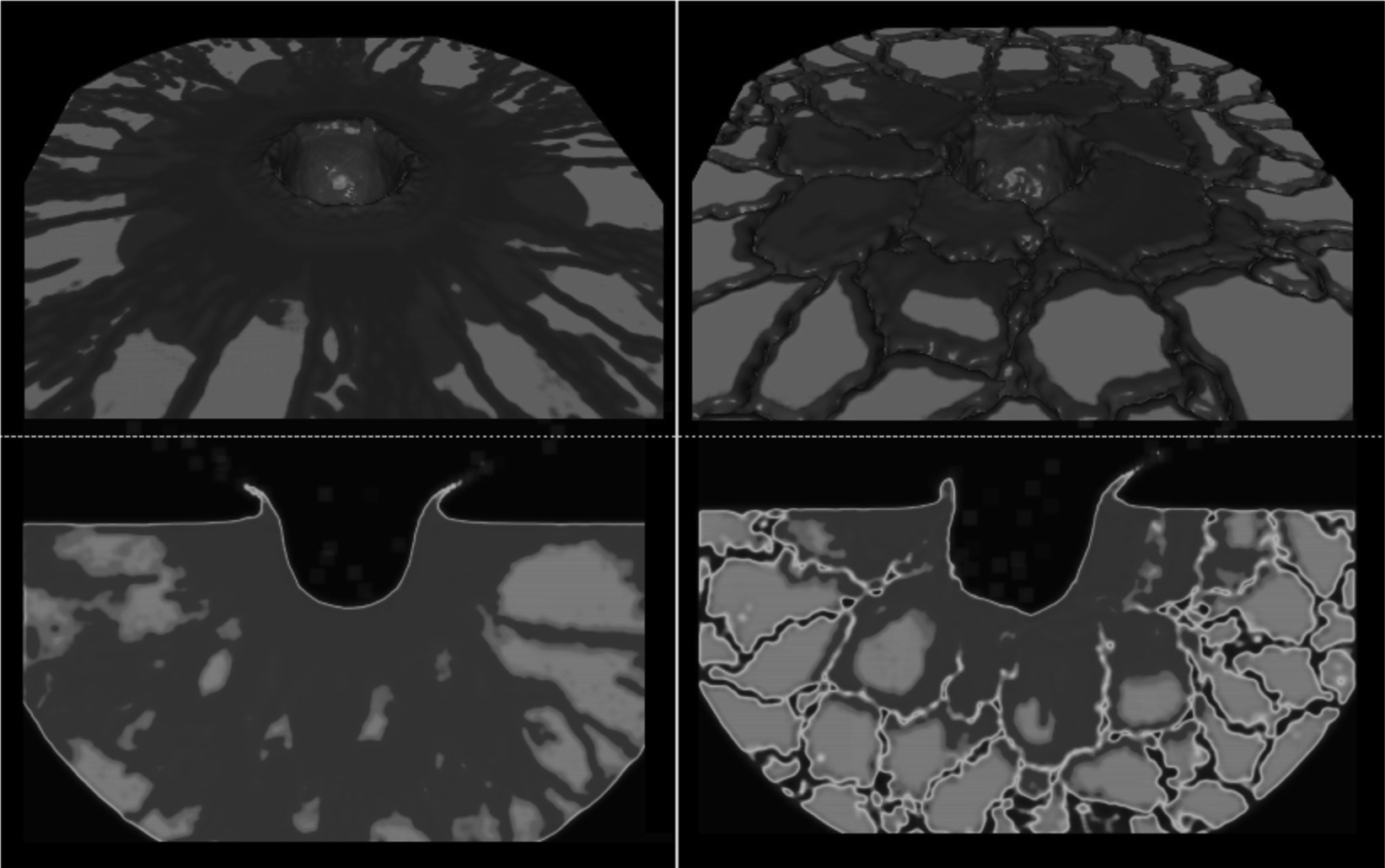}
 \caption{\small Left: damage (dark gray zones) produced by the impact at 10 km/s on a microporous target; the top figure shows the target from above, while the bottom figure shows a vertical slice. Right: same for a  target containing both microporosity and macroporosity. From \emph{Jutzi and Michel} (2014).} \label{figure8}
 \end{figure}
 
In Figure \ref{figure8}, the outcome (in terms of damage) of an impact at 10 km$/$s is shown for the two target structures.

\begin{figure}[t]
 \epsscale{0.9}
 \plotone{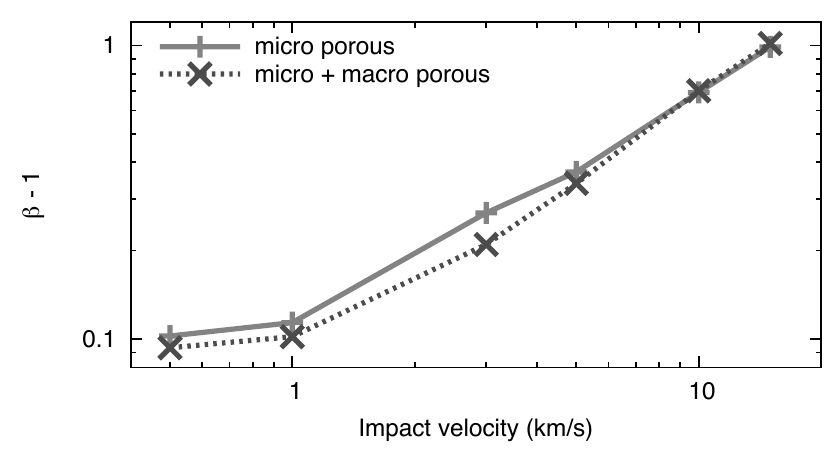}
 \caption{\small Momentum multiplication factor $\beta-1$ as a function of impact velocity for the two considered structures (a: homogeneous microporous, and b: heterogeneous, micro and macroporous).  From \emph{Jutzi and Michel} (2014).}  \label{figure9}
 \end{figure}

In Figure  \ref{figure9}, the results of the calculations of $\beta$ are shown for the considered range of impact velocities (0.5-15 km/s) and the two target structures. At low impact velocities, the amount of momentum transferred is smaller using structure (b) than structure (a). However, these differences disappear at high velocities. This means that for porous targets, inhomogeneities at the scales considered here have negligible effects on the amount of transferred momentum for velocities $\ge$ 5 km$/$s. That confirms the view expressed earlier that the exact form of the modeling of porosity will not matter for any phenomena with scales large compared to the size scale of that porosity.

While the effect of target inhomogeneities on the momentum multiplication factor  $\beta$ appears to be quite small, the effect of various material properties such as the tensile or crushing strength was found to be more significant (\emph{Jutzi and Michel}, 2014).

The results of this study verify that the momentum multiplication factor $\beta$ is small even for very high impact velocities ($\beta < 2$ for $v_{imp} \le15$ km$/$s) in the case of porous targets (with $\sim$ 50\% porosity). This is consistent with scaling laws (section 3) which, in combination with the results of laboratory experiments, predict a small value of $\beta\sim1-2$ for porous materials and comparable impact velocities (see Table 3 in \emph{Holsapple and Housen}, 2012). It is also consistent with the numerical simulations performed by \emph{Holsapple and Housen} (2013a).

However, that is not true for non-porous targets. Values of $\beta$ as large as 5 have been directly measured at impact velocities of $v_{imp} \le5$ km$/$s (\emph{Holsapple and Housen}, 2013a) for rocky targets.  Furthermore, the scaling predicts a marked increase like $U^{0.5} $ for those targets, so at $v_{imp} \le20$ km$/$s those values might reach as much as $\beta \sim10$.  If so, that would significantly facilitate deflections by impacts.

An important question is how to scale strength (tensile strength and crush-curve parameters) of a porous material to larger sizes.  The strength properties of real asteroid materials and their size dependency are not well constrained (e.g., what is the crush-curve of a 300 m  asteroid with 50\% porosity?), while they have a significant effect on the momentum transfer efficiency.

\bigskip
\noindent
\textit{ 5.3. Selective sampling in catastrophic disruptions}
\bigskip

\begin{figure}[t!]
 \epsscale{0.99}
 \plotone{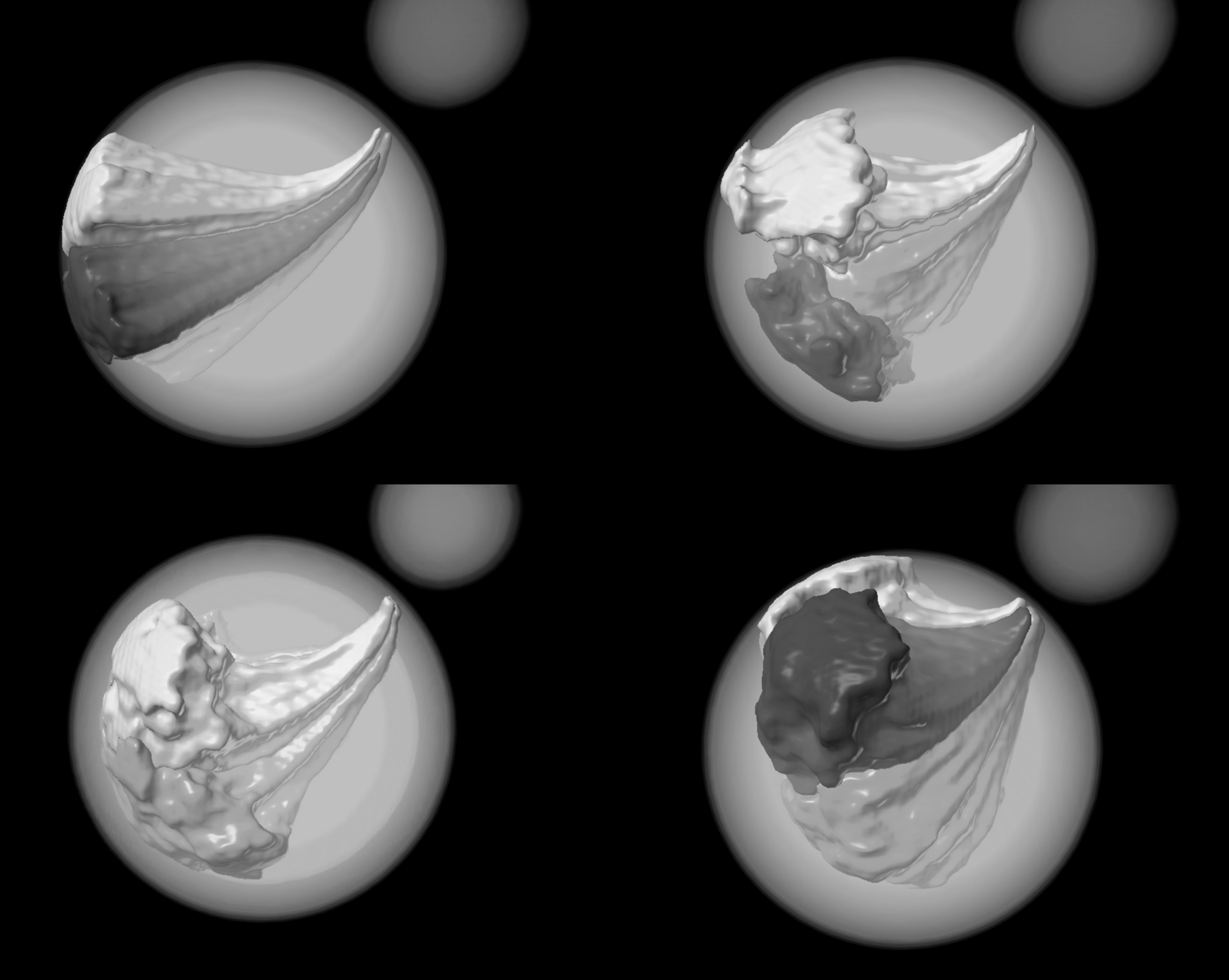}
 \caption{\small Original positions of the material that will reaccumulate and form the largest, second, and third largest fragments (plume-shaped objects shown in different levels of gray). The results for four different internal structures are shown: top left: fully molten; top right: half-molten; bottom left: with a thin molten layer (10\% of the total mass) at 10 km depth (the third largest is hidden behind the visible two largest); bottom right: fully solid. The transparent gray bodies indicate the original target and impactor, respectively. The impactor moves vertically down. In the investigated greatly disruptive regimes, all the material that is in not in the largest fragments (i.e., the largest fraction of the parent body) is blown away, i.e., it will not reaccumulate (or only in very small fragments).   Inspired from \emph{Michel et al.} (2014).} \label{figure10}
 \end{figure}

The catastrophic disruption of a large asteroid as a result of a collision with a smaller projectile and the subsequent reaccumulation of fragments as a result of their mutual gravitational attractions have been simulated numerically during the last decades, which allowed to be reproduced successfully the formation of asteroid families (see \emph{Michel et al.}, this volume). It is generally found that most large bodies formed during a catastrophic disruption consist of aggregates formed by reaccumulation of smaller fragments (e.g. \emph{Michel et al.} 2001). The original location within the parent body of the small pieces that eventually reaccumulate to form the largest offspring of a disruption as a function of the original internal structure of the disrupted asteroid is interesting to determine for several reasons. If reaccumulation is a random process, we expect the particles of a given large fragment to originate from uncorrelated regions within the parent body. Conversely, if the initial velocity field imposed by the fragmentation process determines the reaccumulation phase, the particles belonging to the same fragment should originate from well defined areas in-side the parent body. In addition, the position and extent of these regions provide indications about the mixing occurring as a result of the reaccumulation process.  Motivated by the question of the origin of ureilites, which, in some petrogenetic models, are inferred to have formed at particular depths within their parent body, Michel et al. (2014) started to investigate this problem by simulation the fragmentation (and reaccumulation) of a large body with a diameter fixed at 250 km. This study was performed using a hybrid approach with a SPH shock physics code (\emph{Benz and Asphaug} 1994,1995; \emph{Jutzi et al.} 2008, \emph{Jutzi}, 2014) and the n-body code \emph{pkdgrav} as described in section 4. They considered four kinds of internal structures that may represent the internal structure of a large body in various early stages of the Solar System evolution: fully molten, half molten (i.e., a 26 km-deep outer layer of melt containing half of the mass), solid except a thin molten layer (8 km thick) centered at 10 km depth, and fully solid. The properties of basalt material were assumed for the solid component. They focused on the three largest off spring that had enough reaccumulated pieces to consider. As found by Michel et al. (2004) who considered fully solid and pre-shattered bodies in the context of family formation, they found that the particles that eventually reaccumulate to form the largest reaccumulated bodies retain a memory of their original locations in the parent body. In other words, most particles in each reaccumulated body are clustered from the same original region, even if their reaccumulations take place far away. However, they also found that the extent of the original region varies considerably depending on the internal structure of the parent (Fig. \ref{figure10}). In particular, it seems to shrink with the solidity of the body. Although the covered parameter space in this first study was limited, this sort of investigation can give some constraints on the internal structure of parent bodies of some meteorites. 

\bigskip 
\centerline{\textbf{ 6. CONCLUSIONS}}
\bigskip

Numerical simulations provide an important tool that allow us to probe regimes unreachable by experimental methods. Collisions among asteroids take place in those regimes. To realistically model these events, the combined effects of gravity, strength, porosity as well as shape and structural properties need to be taken into account. Therefore, high fidelity physical models are required, and the modeling asteroid collisions is extremely complex. 

Our knowledge of the asteroid properties which are relevant in terms of impact modeling is still quite limited. Moreover, there are known important shortcomings in the modeling that has been used to date,  although that modeling is improving. It is important to keep that in mind in assessing the meaning of any numerical simulation.

An important step in the development of numerical methods (which includes the implementation of complex material models) is the validation against laboratory experiments. Although  only very briefly  mentioned in this chapter, experimental studies greatly contributed to our understanding of the collisional process at small  laboratory scales and always give a necessary point of reference to test numerical methods. However, in comparisons to experiments it is important to include more than one single scalar value such as the crater size or the fragment mass from one experiment. In order for a numerical model to be a predictive tool, it should be able to reproduce multiple experiments in different regimes without adjusting any parameters. 

Since the publication of Asteroids III a decade ago, there have been major advances in the modeling of asteroid collisions and impact processes. The numerical simulations will undoubtedly become of higher fidelity as our models improve, guided by experimental, theoretical and scaling results. 

\bigskip

\textbf{ Acknowledgements}
M.J. acknowledges the support of the Swiss National Science Foundation through the Ambizione program. K. H. acknowledges support from the NASA grant NNX10AG51G. K.W. acknowledges support through DFG grant WU355/6-1/2. P. M. acknowledges support from the french space agency CNES and the French National Program of Planetology. We thank Gareth Collins and Boris Ivanov for their insightful reviews.
\bigskip

\centerline\textbf{ REFERENCES}
\bigskip
\parskip=0pt
{\small
\baselineskip=11pt

\refs Amsden, A., Ruppel, H., and Hirt, C. (1980) SALE: A simplified ALE computer program for fluid flow at all speeds. Los Alamos National Laboratories Report, LA-8095:101p. Los Alamos, New Mexico: LANL.

\refs Anderson C. E. (1987) An overview of the theory of Hydrocodes. Int. J. Impact Eng. 5, 33-59. 

\refs Artemieva N. and Morgan J. (2009) Modeling the formation of the K-Pg boundary layer. Icarus 201:768-780.
 
\refs Asphaug, E., Ryan, E.V., Zuber, M.T. (2002) Asteroid interiors. In: Bottke, W.F., Cellino, A., Paolicchi, P., Binzel, R.P. (Eds.), Asteroids III. Univ. of Arizona Press, Tucson, pp. 463-484.

\refs Ballouz, R. -L., Richardson, D. C., Michel, P., Schwartz, S. R. (2014a) Rotation-dependent catastrophic disruption of gravitational aggregates. Astrophys. J., 789, 158.

\refs Ballouz, R. -L., Richardson, D. C., Michel, P., Schwartz, S. R., Yu, Y. (2014b) Numerical Simulations of Collisional Disruption of Rotating Gravitational Aggregates: Dependence on Material Properties. Planetary \& Space Sci., accepted.

\refs Barnes, J. and Hut, P. (1986) A Hierarchical O(N log N ) Force-Calculation Algorithm. Nature, 324:446.

\refs Benavidez, P. G. et al. (2012) A comparison between rubble-pile and mono- lithic targets in impact simulations: Application to asteroid satellites and family size distributions. Icarus 219, 57-76.

\refs Benz, W., Cameron, A.G.W. and Melosh, H.J. (1989) The origin of the Moon and the single-impact hypothesis III. Icarus, 81, 113Ð131.
 
\refs Benz, W., Asphaug, E., Jan. (1994) Impact simulations with fracture. I - Method and tests. Icarus 107, 98.  
  
\refs Benz, W. and Asphaug, E. (1995) Simulations of brittle solids using smooth particle hydrodynamics. Computer Physics Communications 87, 253--265

\refs  Benz, W. and Asphaug, E. (1999) Catastrophic disruptions revisited. Icarus 142, 5--20

\refs Berger M.J., Colella P., (1989) Local adaptive mesh refinement for shock hydrodynamics. JCP, 82, 64

\refs Berger M.J., Oliger J., (1984) Adaptive mesh refinement for hyperbolic partial differential equations. JCP, 53, 484

\refs Bray V. J., Collins, G. S., Morgan, J. V., Melosh, H. J., Schenk, P.M. (2014) Hydrocode simulation of Ganymede and Europa cratering trends - How thick is Europa's crust?, Icarus 231, 394-406.

\refs Britt, D.T., Consolmagno, G.J., Merline, W.J., (2006) Small body density and porosity: New data, new insights. Lunar Planet. Sci. XXXVII. Abstract 2214.

\refs Britt, D.T., Yeomans, D., Housen, K., Consolmagno, G.J (2002) Asteroid Density, Porosity, and Structure. In: Bottke, W.F., Cellino, A., Paolicchi, P., Binzel, R.P. (Eds.), Asteroids III. Univ. of Arizona Press, Tucson, pp. 485-500.

\refs Canup, R.M. and Asphaug, E. (2001) Origin of the Moon in a giant impact near the end of the EarthÕs formation. Nature, 412, 708Ð712.

\refs Canup R. M., Barr A. C., Crawford D.A.( 2013) Lunar-forming impacts: high-resolution SPH and AMR-CTH simulations. Icarus 222, 200-219. 
 
\refs Carroll, M.M., Holt, A.C., (1972) Suggested modification of the $P$ - $\alpha$ model for porous materials. 
J. Appl. Phys. 43, 759-761.

\refs 	Chapman, C. R., McKinnon, W. B. (1986) Cratering of planetary satellites. Satellites, p. 492 - 580.

  
\refs Collins, G. S., Melosh, H. J., Ivanov, B. A (2004) Modeling damage and deformation in impact simulations. Meteorit. Planet. Sci. 39, 217--231  

\refs Collins, G. S., Melosh, H. J., W\"unnemann, K. (2011) Improvements to the $\epsilon$-$\alpha$ porous compaction model for simulating impacts into high-porosity solar system objects. Int. J. Imp. Eng. 38, 434-439.

\refs Collins, G. S., W\"unnemann, K., Artemieva, N., Pierazzo, E. (2013) Numerical modelling of impact processes, in Impact Cratering: Processes and Products 254--270 (Wiley-Blackwell). ISBN: 9781405198295  

\refs Collins, G. S. (2014) Numerical simulations of impact crater formation with dilatancy. Journal of Geophysical Research (Planets), in press, doi:10.1002/2014JE004708

\refs Cheng, A.F. (2012)  Asteroid deflection by spacecraft impact. Asteroids, Comets, Meteors 2012, Proceedings. LPI Contribution No. 1667 


\refs Cremonese, G.,  E. Martellato, F. Marzari, E. Kuhrt, F. Scholten, F. Preusker, K. WŸnnemann, P. Borin, M. Massironi, E. Simioni, W. (2012)
Hydrocode simulations of the largest crater on the asteroid Lutetia
Planetary and Space Science, 66, pp. 147Ð154

\refs Davison, T. M., Collins, G. S.,, Ciesla, F. J. (2010) Numerical modelling of heating in porous planetesimal collisions. Icarus, 208(1), 468Ð481.

\refs Davison, T. M., Collins, G. S., Elbeshausen, D., WŸnnemann, K., Kearsley, A. (2011) Numerical modeling of oblique hypervelocity impacts on strong ductile targets. Meteoritics \& Planetary Science, 46(10), 1510Ð1524.

\refs Davison, T. M., Ciesla, F. J., Collins, G. S. (2012) Post-Impact Thermal Evolution of Porous Planetesimals. Geochimica et Cosmochimica Acta 95, 252-269.

\refs Dienes, J.K., Walsh, J.M., (1970) Theory of impact: Some general principles and the method of Eulerian codes, in High Velocity Impact Phenomena, edited by R. Kinslow, pp. 45-104, Academic Press, New York.

\refs Elbeshausen D., W\"unnemann K., Collins G. S. (2009) Scaling of oblique impacts in frictional targets: Implications for crater size and formation mechanisms. Icarus  doi:10.1016/ j.icarus.2009.07.018.

\refs Elbeshausen D., W\"unnemann K. (2010) iSALE-3D: A three-dimensional, multi-material, multi-rheology hydrocode and its applications to large-scale geodynamic processes. In: Proceedings of 11th Hypervelocity Impact Symposium, 287-301. 

\refs Elbeshausen, D., WŸnnemann, K., Collins, G. S. (2013). The transition from circular to elliptical impact craters. Journal of Geophysical Research (Planets), 118(11), 2295Ð2309. doi:10.1002/2013JE004477

\refs Fujiwara, A. et al., (2006) The rubble-pile asteroid Itokawa as observed by Hayabusa. Science 312, 1330-1334.

\refs Flynn, G.  J., Durda, D. D., Sandel, L.  E.,  Kreft, J.  W., Strait, M. M. (2008) Dust production from the hypervelocity impact disruption of the Murchison hydrous CM2 meteorite: Implications for the disruption of hydrous asteroids and the production of interplanetary dust. Planet. Space Sci., 57, 119-126.

\refs Gault, D.E., (1973) The Moon 6, 32Ð44.

\refs Geretshauser, R.J., Speith, R. and Kley, W.  (2011) Collisions of inhomogeneous pre-planetesimals. A\&A 536, A104 

\refs Gingold, R.A., Monaghan, J.J. (1977) Smoothed particle hydrodynamics - Theory and application to non-spherical stars. Royal Astronomical Society, 181, 375.

\refs Grady, D.E., Kipp, M.E., (1980). Continuum modelling of explosive 
fracture in oil shale. Int. J. Rock Mech. Min. Sci. \& Geomech. Abstr. 17, 147--157.

\refs G\"uldemeister, Ni. W\"unnemann, K., Durr, N., Hiermaier, S. (2013) Propagation of impact-induced shock waves in porous sandstone using mesoscale modeling. Meteoritics \& Planetary Science, Volume 48, Issue 1, pp. 115-133.

\refs Harrison, K., and R. Grimm (2003), Rheological constraints on Martian landslides, Icarus, 163, 347Ð362).

\refs Herrmann, W., (1969) Constitutive equation for the dynamic compaction of ductile porous materials. 
J. Appl. Phys., 40, 2490-2499.  

\refs Hiraoka, K., Hakura, S., Nakamura, A. M., Suzuki, A., Hasegawa, S. (2011) Impact Cratering Experiments on Porous Sintered Targets of Different Strengths. Meteoritics and Planetary Science Supplement, 74, 5363.

\refs Hirt, C. W.; Amsden, A. A.; Cook, J. L. (1974) An Arbitrary Lagrangian-Eulerian Computing Method for All Flow Speeds. Journal of Computational Physics, Volume 14, Issue 3, p. 227-253.

\refs Hoek E. and Brown E.T. (1980) "Empirical strength criterion for rock masses". J. Geotechnical Engineering Division ASCE: 1013Ð1025.

\refs Holsapple, K.A., (1993) The scaling of impact processes in planetary sciences. Annu. Rev. Earth Planet. Sci. 21, 333Ð373.

\refs Holsapple (2004a) From Simple to Complex Craters: The Mechanics of Late-time Crater Adjustments. 35th Lunar and Planetary Science Conference, p. 1937.

\refs Holsapple (2004b) Mitigation of hazardous comets and asteroids. Edited by M. Belton, T. H. Morgan, N. Samarasinha, and D. K. Yeomans.  Cambridge University Press, Cambridge, UK, 2004, p.113.

\refs Holsapple, K.A. (2008) Models of porosity for impact events. In: Proceedings of the 39th LPSC, Houston, TX, USA.
    
\refs Holsapple, K.A. (2009) On the ``strength" of the small bodies of the solar system: A review of strength theories and their implementation for analyses of impact disruptions. Planetary and Space Science 57, 127--141

\refs Holsapple, K.A. (2013) Modeling granular material flows: The angle of repose, fluidization and the cliff collapse problem. Planetary and Space Science 82-83, 11--26

\refs Holsapple, K.A., Giblin, I., Housen, K., Nakamura, A., Ryan, E., (2002) Asteroid impacts: Laboratory experiments and scaling laws.  In: Bottke,W.F., Cellino, A., Paolicchi, P., Binzel, R.P. (Eds.), Asteroids III. Univ. of Arizona
Press, Tucson, pp. 443.

\refs Holsapple, K.A. and Housen, K.R. (2012) Momentum transfer in asteroid impacts. I. Theory and scaling.  Icarus 221, 875--887

\refs Holsapple, K.A. and Housen, K.R. (2013a) Mitigation by Impacts: Theory, experiments, and code calculations. In Proceedings of the IAA Planetary Defense Conference 2013, Flagstaff, USA.

\refs Holsapple, K.A. and Housen, K.R. (2013b) The Third Regime of Cratering: Spall Craters. 44th Lunar and Planetary Science Conference, vol 44, p. 2733.

\refs Holsapple and Schmidt (1982) Journal of Geophysical Research, vol. 87, Mar. 10, 1982.

\refs Housen, K. R. and Holsapple, K. A. (1990) On the fragmentation of asteroids and planetary satellites. Icarus, 84, 226.



\refs Housen, K.R., Holsapple, K.A. (2011) Ejecta from impact craters. Icarus 211, 856--875.

\refs Housen, K.R. and Sweet, W. J. (2013) Experimental simulation of  large-scale impacts on porous asteroids. 44th Lunar and Planetary Science Conference, p. 1993.

\refs Ivanov B. A., Deniem D., and Neukum G. (1997) Implementation of dynamic strength models into 2D hydrocodes: Applications for atmospheric breakup and impact cratering. Int. J. Impact Eng. 20, 411-430.
 
\refs Jop, P., Y. Forterre, and O. Pouliquen (2006) A constitutive law for dense granular flows, Nature, 441, 727--730. 
  
\refs Jutzi, M., Benz, W., Michel, P. (2008) Numerical simulations of impacts involving porous bodies. I. Implementing sub-resolution porosity in a 3D SPH hydrocode. Icarus 198, 242--255  
  
\refs  Jutzi, M., Michel, P., Hiraoka, K, Nakamura, A.M., Benz, W. (2009) Numerical simulations of impacts involving porous bodies. II. Comparison with laboratory experiments. Icarus 201, 802--813  

\refs Jutzi, M., Michel, P., Benz, W., Richardson, D.C. (2010) Fragment properties at the catastrophic disruption threshold: The effect of the parent bodys internal structure. Icarus 207, 54-65

\refs  Jutzi, M., Asphaug, E., Gillet, P., Barrat, J-A., Benz, W. (2013) The structure of the asteroid 4 Vesta as revealed by models of planet-scale collisions. Nature 494, 207--210

\refs  Jutzi, M., Michel, P. (2014) Hypervelocity impacts on asteroids and momentum transfer I. Numerical simulations using porous targets. Icarus 229, 247--253 

\refs Jutzi, M. (2014) SPH calculations of asteroid disruptions: the role of pressure dependent failure models. Planetary and Space Science, in press. 

\refs Kenkmann, T., WŸnnemann, K., Deutsch, A., Poelchau, M. H., SchŠfer, F., Thoma, K (2011) Impact cratering in sandstone: The MEMIN pilot study on the effect of pore water. Meteoritics \& Planetary Science, Volume 46, Issue 6, pp. 890-902. 

\refs Kenkmann, T., Deutsch, A., Thoma, K., Poelchau, M. (2013) The MEMIN research unit: Experimental impact cratering. Meteoritics \& Planetary Science, Volume 48, Issue 1, pp. 1-2.

\refs Kimberley, J., Ramesh, K.T. (2011) The dynamic strength of an ordinary chondrite. Meteoritics \& Planetary Science 46, Nr 11, 1653--1669.

\refs Kaplinger, B., Premaratne, P., Setzer, Ch. and Wie, B. (2013) GPU Accelerated 3-D Modeling and Simulation of a Blended Kinetic Impact and Nuclear Subsurface Explosion. Planetary Defense Conference 2013 IAA-PDC13-04-06. 

\refs Kowitz, A., Schmitt, R. T., Uwe R.W., Hornemann, Ul. (2013) The first MEMIN shock recovery experiments at low shock pressure (5-12.5 GPa) with dry, porous sandstone. Meteoritics \& Planetary Science, Volume 48, Issue 1, pp. 99-114.

\refs Krohn, K., et al., Asymmetric craters on Vesta: Impact on sloping surfaces. Planetary and Space Science (2014),
http://dx.doi.org/10.1016/j.pss.2014.04.011.

\refs Leinhardt, Z. M., Stewart, S. T. (2009). Full numerical simulations of catastrophic small body collisions. Icarus, 199, 542. doi:10.1016/j.icarus.2008.09.013

\refs Leinhardt, Z.M., Stewart, S.T. (2012) Collisions between Gravity-dominated Bodies. I. Outcome Regimes and Scaling Laws.  The Astrophysical Journal, 745,  Issue 1, article id. 79, 27 pp.

\refs Leliwa-Kopystynski, J. Arakawa, M. (2014) Impacts experiments onto heterogeneous targets simulating impact breccia: Implications for impact strength of asteroids and formation of the asteroid families. Icarus, 235, 147-155. 
  

\refs Lubarda, V. and Krajcinovic, D. (1993) Damage Tensors and the Crack Density Distribution, Int. J. Solids Structures, Vol. 30, pp. 2859-2877

\refs Lucas, A., and A. Mangeney (2007) Mobility and topographic effects for large Valles Marineris landslides on Mars, Geophys. Res. Lett., 34, L10201, doi:10.1029/2007GL029835

\refs Marcus, R.A.,  Stewart, S.T., Sasselov, D., Hernquist, L. (2009) Collisional stripping and disruption of super-Earths. Astrophys. Lett. 700, L118 doi:10.1088/0004-637X/700/2/L118

\refs Maxwell (1977). In Impact and explosion cratering. Planetary and terrestrial implications, p. 1003 - 1008.


\refs McGlaun, J.M., Thompson, S.L., Elrick, M.G. (1990) CTH: A 3-dimensional shock-wave physics code.  Int. J. Imp. Eng., 10, 351-360

\refs Melosh, H.J., (1979) Acoustic Fluidization: A New Geologic Process? JGR 84, 7513-7520.

\refs Melosh, H.J. (2007) A hydrocode equation of state for SiO2. Meteoritics and Planetary Science, 42, 2079--2098.

\refs Melosh, H. J. and Ivanov, B. A. (1999) Impact crater collapse. Annu. Rev. Earth Planet. Sci. 27,
385--415.

\refs Melosh H. J., Ryan E. V., and Asphaug E. (1992) Dynamic Fragmentation in Impacts: Hydrocode Simulation of Laboratory Impacts. J. Geophys. Res. 97(E9), 14735-14759.

\refs Michel, P., Benz, W., Tanga, P., Richardson, D.C. (2001). Collisions and gravitational reaccumulation: Forming asteroid families and satellites. Science 294, 1696--1700.

\refs Michel, P., Benz, W. and Richardson, D. C. (2004) Disruption of pre-shattered parent bodies. Icarus, 168, 420-432.

\refs Michel, P., Jutzi, M., Richardson, D. C., Goodrich, C. A., Hartmann, W. K., OÕBrien, D. P. (2014) Selective sampling during catastrophic disruption: Mapping the location of reaccumulated fragments in the original parent body. Planet. Space Sci., in press.

\refs Monaghan, J.J. (2012) Smoothed Particle Hydrodynamics and Its Diverse Applications. Annu. Rev. Fluid Mech. 2012. 44:323--46. 

\refs Nakamura A., Yamane, F.,  Okamoto, T., Takasawa S. (2014) Size dependence of the disruption threshold: laboratory examination of millimeter-centimeter porous targets. Planetary and Space Science, in press.

\refs Okamoto, C., Arakawa, M. (2008) Experimental study on the impact fragmentation of core mantle bodies: Implications for collisional disruption of rocky planetesimals with sintered core covered with porous mantle. Icarus, 197, 627-637.

\refs Okamoto, C., Arakawa, M. (2009) Experimental study on the collisional disruption of porous gypsum spheres. Meteoritics and Planetary Science, 44, 1947-1954.

\refs Owen, J.M. (2010) ASPH modeling of material damage and failure, Proceedings of the fifth International SPHERIC workshop, 297-304

\refs Pierazzo E., Artemieva N., Asphaug E., Baldwin E. C., Cazamias J., Coker R., Collins G. S., Crawford D., Elbeshausen D., Holsapple K. A., Housen K. R., Korycansky D. G., and W\"unnemann K. (2008) Validation of numerical codes for impact and explosion cratering. Meteoritics and Planetary Science 43(12), 1917-1938.



\refs Reufer, A., Meier, M. M. M., Benz, W., Wieler, R. (2012) A hit-and-run giant impact scenario. Icarus, 221(1), 296Ð299. doi:10.1016/j.icarus.2012.07.021

\refs Richardson, D. C., T. Quinn, J. Stadel, and G. Lake (2000) Direct large-scale N-body simulations of planetesimal dynamics. Icarus 143, 45Ð 59.

\refs Richardson, D.C., Leinhardt, Z.M., Melosh, H.J., Bottke Jr., W.F., Asphaug, E., (2002) Gravitational aggregates: Evidence and evolution. In: Bottke, W.F., Cellino, A., Paolicchi, P., Binzel, R.P. (Eds.), Asteroids III. Univ. of Arizona Press, Tucson, pp. 501-515.

\refs Richardson, D. C., Walsh, K., Murdoch, N., Michel, P. (2011) Numerical simulations of granular dynamics. I. Method and tests. Icarus, 212, 427-437.

\refs P. Sanchez and D.J. Scheeres (2011) Simulating asteroid rubble piles with a self-gravitating soft-sphere distinct element method model. The Astrophysical Journal, 727:120.

\refs Sanchez, P., Scheeres, D. J. (2014) The strength of regolith and rubble pile asteroids. Meteoritics and Planetary Science, 49, 788-811. 

\refs Schmidt, R.M. (1977) A centrifuge cratering experiment: development of a gravity-scaled yield parameter. Impact and Explosive Cratering. New York: Pergamon. p. 1261-78

\refs Schmidt, R.M., Holsapple, K.A., (1980) Theory and experiments on centrifuge cratering. J. Geophys. REs. 85, 235-52.


\refs Senft. L.E., Stewart, S.T. (2007) Modeling impact cratering in layered surfaces. JOURNAL OF GEOPHYSICAL RESEARCH, VOL. 112, E11002, doi:10.1029/2007JE002894, 2007.


\refs Setoh, M., Nakamura, A.M., Michel, P., Hiraoka, K, Yamashita, Y., Hasegawa, S., Onose, N., Okuidara, K. (2010) High and low-velocity impact experiments on porous sintered glass bead targets of different compressive strengths: outcome sensitivity and scaling. Icarus 205, 702-711.

\refs Shapiro P., H. Martel, J. Villumsen, and J. Owen (1996) Astron. J. Suppl. Ser., vol. 103, pp. 269--330.

\refs Shuvalov V. V. (1999) Multi-dimensional hydrodynamic code SOVA for interfacial flows: Application to the thermal layer effect, Shock Waves 9:381-390.

\refs Springel, V. (2005) The cosmological simulation code GADGET-2. Mon. Not. R. Astron. Soc. 364, 1105. doi:10.1111/j.1365-2966.2005.09655.x

\refs Stewart, S.T. and Leinhardt, Z.M. (2009) Velocity-dependent catastrophic disruption criteria for planetesimals. The Astrophysical Journal, 691:L133L137

\refs St\"offler, D., Artemieva, N.A., Pierazzo, E. (2002) Modeling the Ries-Steinheim impact event and the formation of the moldavite strewn field. Meteoritics \& Planetary Science 37 1893-1907.

\refs Schwartz, S. R., Richardson, D. C., Michel, P. (2012) An Implementation of the Soft-Sphere Discrete Element Method in a High-Performance Parallel Gravity Tree-Code. Granular Matter, DOI 10.1007/s10035-012-0346-z.

\refs Schwartz, S. R., Michel, P., Richardson, D. C. (2013) Numerically simulating impact disruptions of cohesive glass bead agglomerates using the soft-sphere discrete element method. Icarus, 226, 67-76.

\refs Thompson, S.L. and Lauson, H.S. (1972) Improvements in the chart- D radiation hydrodynamic code III: revised analytical equation of state. Report SC-RR-710714, Sandia National Laboratories, Albuquerque, NM.

\refs Tillotson, J.H. (1962) Metallic equations of state for hypervelocity
impact. General Atomic Report GA-3216, July 1962.

\refs Walker, J. D., Chocron, S., Durda, D. D., Grosch, D. J., Movshovitz, N., Richardson, D.C., Asphaug, E. (2013) Momentum enhancement from aluminum striking granite and the scale size effect. International Journal of Impact Engineering, 56, 12-18.

\refs Weibull, W., (1939) A statistical theory of strength of materials. R. Swedish Inst. Eng. Res., 1--45.

\refs W\"unnemann K., Ivanov B. A. (2003) Numerical modelling of impact crater depth-diameter dependence in an acoustically fluidized target. Planetary and Space Science, 51, 831-845.

\refs W\"unnemann, K., Collins, G.S. and Melosh, H.J. (2006) A strain-based porosity model for use in hydrocode simulations of impacts and implications for transient crater growth in porous targets, Icarus, 180, 514--527.

\refs W\"unnemann K., Nowka D., Collins G.S., Elbeshausen D., Bierhaus M. (2010) Scaling of impact crater formation on planetary surfaces – insights from numerical modeling. In: Proceedings of 11th Hypervelocity Impact Symposium, 1-13.

\refs Yasui, M., Arakawa, M. (2011) Impact experiments of porous gypsum-glass bead mixtures simulating parent bodies of ordinary chondrites: Implications for re-accumulation processes related to rubble-pile formation. Icarus, 214, p. 754-765.

\refs Yeomans, D.K., and 12 colleagues (1997) Estimating the mass of Asteroid 253 Mathilde from tracking data during the NEAR flyby. Science 278, 2106-2109.

\end{document}